\documentclass[11pt]{article}
\usepackage{latexsym}
\usepackage{amssymb}
\usepackage{graphics}
\usepackage{epsfig}
\usepackage{rotating}
\usepackage{amsfonts}
\usepackage{amsmath}
\usepackage{hyperref}
\hypersetup{
	colorlinks=true,
	linkcolor=blue,
	filecolor=magenta,
	urlcolor=cyan,
}
\begin{document}

\begin{titlepage}
\begin{flushright}
IRMP-CP3-23-18\\
\end{flushright}

\vspace{5pt}

\begin{center}

{\Large\bf External gauge field coupled quantum dynamics:}\\

\vspace{7pt}

{\Large\bf gauge choices, Heisenberg algebra representations}\\

\vspace{7pt}

{\Large\bf and gauge invariance in general,}\\

\vspace{7pt}

{\Large\bf and the Landau problem in particular}\\

\vspace{60pt}

Jan Govaerts$^{a,b,}$\footnote{Fellow of the Stellenbosch Institute for Advanced Study (STIAS), Stellenbosch,
Republic of South Africa}

\vspace{30pt}

$^{a}${\sl Centre for Cosmology, Particle Physics and Phenomenology (CP3),\\
Institut de Recherche en Math\'ematique et Physique (IRMP),\\
Universit\'e catholique de Louvain (UCLouvain),\\
2, Chemin du Cyclotron, B-1348 Louvain-la-Neuve, Belgium}\\
E-mail: {\em Jan.Govaerts@uclouvain.be}\\
ORCID: {\tt http://orcid.org/0000-0002-8430-5180}\\
\vspace{15pt}
$^{b}${\sl International Chair in Mathematical Physics and Applications (ICMPA--UNESCO Chair)\\
University of Abomey-Calavi, 072 B.P. 50, Cotonou, Republic of Benin}\\

\vspace{10pt}

%\today

\vspace{20pt}

\begin{abstract}
\noindent
Even though its classical equations of motion are then left invariant, when an action is redefined by an additive total derivative or divergence term
(in time, in the case of a mechanical system) such a transformation induces nontrivial consequences for the system's canonical phase space
formulation. This is even more true and then in more subtle ways for the canonically quantised dynamics, with in particular an induced transformation in the unitary
configuration space representation of the Heisenberg algebra being used for the quantum system. When coupled to a background gauge field,
such considerations become crucial for a proper understanding of the consequences for the system's quantum dynamics
of gauge transformations of that classical external background gauge field, while under
such transformations the system's degrees of freedom, abstract quantum states and quantum dynamics are certainly strictly invariant.
After a detailed analysis of these different points in a general context, these are then illustrated
specifically in the case of the quantum Landau problem with its classical external background magnetic vector potential for which the most general
possible parametrised gauge choice is implemented herein. The latter discussion aims as well to clarify some perplexing statements in the literature
regarding the status of gauge choices to be made for the magnetic vector potential for that quantum system.
The role of the global space-time symmetries of the Landau problem and their gauge invariant Noether charges
is then also emphasized.

\end{abstract}

\end{center}

\end{titlepage}

\setcounter{footnote}{0}

\section{Introduction}
\label{Intro}

The celebrated Landau problem considers the dynamics of an electrically charged nonrelativistic point particle of mass $m$ and charge $q$, of which the motion
restricted to a two dimensional plane is subjected to a static and uniform magnetic field $\vec{B}$ perpendicular to that plane.

Given an inertial frame of right-handed orthonormal basis $\{\hat{e}_1,\hat{e}_2,\hat{e}_3\}$ with its origin fixed in that plane and the vectors
$\{\hat{e}_1,\hat{e}_2\}$ aligned within that plane, so that $\hat{e}_1\times\hat{e}_2=\hat{e}_3$
and $\vec{B}=B\,\hat{e}_3$, the particle's position vector $\vec{x}(t)=\sum_{i=1}^2 x_i(t)\,\hat{e}_i$
then evolves according to the Lorentz force equation,
\begin{equation}
m\frac{d^2\vec{x}(t)}{dt^2}=q\,\frac{d\vec{x}(t)}{dt}\times\vec{B}.
\label{eq:Newton1}
\end{equation}
Equivalently in terms of the components $x_i(t)$ ($i,j=1,2$), one has (the summation convention over repeated indices in products is implicit throughout),
\begin{equation}
m\ddot{x}_i=qB\,\epsilon_{ij}\,\dot{x}_j,\quad
\dot{p}_i=qB\,\epsilon_{ij}\dot{x}_j=\frac{qB}{m}\,\epsilon_{ij}\,p_j=s\,\omega_c\,\epsilon_{ij}\,p_j,\quad
p_i(t) \equiv m\dot{x}_i(t),\quad
\dot{}\equiv \frac{d}{dt},
\end{equation}
with $\epsilon_{ij}=-\epsilon_{ji}$ and $\epsilon_{12}=+1$, while $\omega_c \equiv |qB|/m$ defines the cyclotron frequency and $s={\rm sgn}(qB)=\pm 1$
denotes the sign of the product $qB$, which also determines the rotation orientation for the closed circular trajectories of the point charge.

The general solution for these classical circular trajectories may be given (for instance) in the form,
\begin{eqnarray}
\vec{x}(t) &=& \vec{x}_c+\frac{1}{\omega_c}\sqrt{\frac{2E}{m}}\left(\cos\omega_c (t-t_0)\,\hat{e}_1\,-\,s\,\sin\omega_c(t-t_0)\,\hat{e}_2\right), \nonumber \\
\vec{p}(t) &=& -\sqrt{2mE}\left(\sin\omega_c(t-t_0)\,\hat{e}_1\,+\,s\,\cos\omega_c(t-t_0)\,\hat{e}_2\right),
\end{eqnarray}
with the necessary four integration constants corresponding, in this representation, to the position vector of the magnetic or guiding centre, $\vec{x}_c=x_{ci}\hat{e}_i$,
the circle radius amplitude factor $(\sqrt{2E/m})/\omega_c$ ($E\ge 0$), and the phase factor $(-\omega_c t_0)$ for time dependency.

The \textcolor{black}{vector} equation of motion (\ref{eq:Newton1}) being obviously invariant (or covariant) under continuous and time independent translations
in time, translations in space,
and rotations in space about any given point $\vec{x}_0=x_{0i}\hat{e}_i$, this dynamics is characterised by four conserved quantities or Noether charges,
denoted hereafter $E$, $\vec{T}=T_i\hat{e}_i$ and $M_3$, respectively. These observables extend in the presence of the given magnetic field $\vec{B}$,
the ordinary conserved kinetic energy $\vec{p}\,^2/(2m)$, velocity (or kinetic, or mechanical) momentum $\vec{p}=m\dot{\vec{x}}$,
and orbital angular-momentum $\vec{L}=(\vec{x}-\vec{x}_0)\times\vec{p}=L_3\hat{e}_3$ (relative to $\vec{x}_0$) of a free particle  in the plane, respectively.
As is always the case for Noether charges, when considered within the canonical first-order Hamiltonian formulation
these conserved quantities are also the generators of the corresponding symmetry transformations \textcolor{black}{(see for example Ref.\cite{Gov1})},
while \textcolor{black}{since they are} independent of time as phase-space observables they also possess vanishing Poisson brackets with the Hamiltonian \cite{Gov1,Victor,Daddy}.
In explicit form, one has for the Landau problem, as may be directly read off the equation of motion~(\ref{eq:Newton1}),
\begin{equation}
E=\frac{1}{2m}\vec{p}\,^2,\qquad
T_i=p_i-qB\epsilon_{ij} (x_j-x_{0j}),\qquad
M_3=\epsilon_{ij}(x_i-x_{0i}) p_j + \frac{1}{2}qB(\vec{x}-\vec{x}_0)^2,
\end{equation}
where the arbitrary position vector $\vec{x}_0=x_{0i}\hat{e}_i$ is introduced for the sake of generality (it also plays the role of a new origin in the plane),
while of course $\vec{p}=m\dot{\vec{x}}$. In contradistinction to the conserved energy $E$ which retains its usual \textcolor{black}{purely kinetic} expression
(since a magnetic field
with its Lorentz force does not work), the three conserved quantities $\vec{T}=T_i \hat{e}_i$ and $M_3$ indeed combine the usual corresponding conserved
observables for a free particle with a direct contribution from the magnetic field $\vec{B}$ as well. Therefore one could refer to these as the conserved
total momentum and total angular-momentum of the system, inclusive of the contributions from the magnetic field, respectively.

Note that these four conserved quantities are not independent, since they obey the general relation
\begin{equation}
\vec{T}\,^2=2mE + 2 qB M_3 = 2mE + 2s m\omega_c M_3,\qquad
M_3=-\frac{s}{\omega_c}\left(E-\frac{1}{2m}\vec{T}\,^2\right)=-\frac{m}{qB}E + \frac{1}{2qB}\vec{T}\,^2.
\label{eq:Relation}
\end{equation}

A simple substitution of the general solution for these quantities finds that the amplitude parameter $E\ge 0$ of course coincides with the value
for the system's conserved energy, while the other three conserved observables take the following values,
\begin{equation}
\vec{T}=-q\left(\vec{x}_c-\vec{x}_0\right)\times\vec{B},\qquad
T_i=-qB\epsilon_{ij}(x_{cj}-x_{0j}),\qquad
M_3=-s\frac{E}{\omega_c}+\frac{1}{2}qB\left(\vec{x}_c - \vec{x}_0\right)^2.
\end{equation}
In particular, in units of $qB$, the components of the dual spatial translation generator $\vec{T}\times\hat{e}_3$ measure the position
of the magnetic centre (relative to $\vec{x}_0$; \textcolor{black}{note that} $\vec{T}$ is defined up to an arbitrary constant value anyway), since
\begin{equation}
\vec{x}_c=\vec{x}_0+\frac{1}{qB}\vec{T}\times\hat{e}_3,\qquad
x_{ci}=x_{0i}+\frac{1}{qB}\epsilon_{ij} T_j=x_{0i}+\frac{s}{m\omega_c}\epsilon_{ij} T_j.
\end{equation}
This observation completes the physical interpretation of all four conserved quantities generating the global space-time symmetries of the Landau problem,
and shows how these Noether charges determine the values of the integration constants for the general solution to its dynamics, up to an arbitrary phase factor
which is related to its invariance under constant translations in time (namely the value of $(-\omega_c t_0)$).

When it comes to the quantum dynamics of the Landau problem, subtleties arise because of the need, within a space-time local Hamiltonian formulation
and its canonical quantisation deriving from a local variational principle, of describing the coupling to the external background magnetic field in terms of
a vector potential $\vec{A}(\vec{x}\,)$ such that $\vec{\nabla}\times\vec{A}(\vec{x}\,)=\vec{B}$, but not directly in terms of the magnetic field itself,
while \textcolor{black}{in contradistinction} the magnetic field alone is involved in the above classical dynamics and its conserved charges.
Of course the choice of vector potential is correlated
with the freedom of effecting local gauge transformations in the form of
\begin{equation}
\vec{A}(\vec{x}\,) \rightarrow \vec{A}'(\vec{x}\,)=\vec{A}(\vec{x})+\vec{\nabla}\varphi(\vec{x}\,),\qquad \vec{\nabla}\times\vec{A}'(\vec{x}\,)=\vec{B},
\end{equation}
where $\varphi(\vec{x}\,)$ is an arbitrary scalar function in the plane (which in the present case is time independent, since no extra background
electric field is considered). Yet the quantum physics of the system, to be observed through its \textcolor{black}{abstract} quantum observables and
their matrix elements for \textcolor{black}{abstract} quantum states,
should remain independent of the choice of vector potential, namely it should be invariant under such gauge transformations
{\it of the background field} $\vec{A}(\vec{x}\,)$, as is \textcolor{black}{this physics already at the classical level in a trivial manner indeed}.

\textcolor{black}{Note well that such gauge transformations solely act on the external classical 
background gauge field, while the system's degrees of freedom remain strictly
invariant since they are certainly not subjected to any such transformation. Consequently its purely abstract quantum states should likewise be strictly gauge invariant.
Yet for reasons to be revisited hereafter, {\it their configuration space wave function representations} transform contravariantly under such gauge transformations.}

Aiming to address these subtleties and dispel some confusion that may have arisen regarding such properties, these issues have been discussed and revisited
over the years in a (large) number of publications, the most recent being that of Ref.\cite{Waka1} (which includes a fair collection of references
to the relevant literature \textcolor{black}{that the reader is referred to without reproducing it in the present bibliography}).
Yet rather than dispelling some confusion, sometimes such publications have added to it. It may thus seem timely to address
these issues anew, by emphasizing as well some additional but implicit features in relation to different unitary representations of the Heisenberg algebra,
that may have concurred to produce some of the \textcolor{black}{confusion. In particular in the own words of the authors of Ref.\cite{Waka1}, they view
that situation indeed as perplexing, hence their most recent publication whose ambition is to properly clarify the existing confusion. Unfortunately,
for reasons addressed in detail in the present work,
that aim is not achieved, because of ill-conceived conclusions and statements that affect the discussion of Ref.\cite{Waka1} and add to the confusion.}

\textcolor{black}{In view of such considerations} the discussion \textcolor{black}{hereafter}
is structured as follows. \textcolor{black}{Section~\ref{Sect2} considers, first in a general context independently of the Landau problem,}
the different consequences that result, whether for the
classical or the quantised canonical formulation of a dynamics, from a redefinition of the classical action by an additive surface or divergence term,
or total derivative contribution, specifically a total derivative surface term in time for a mechanical system. Even though the classical equations of motion
are then left invariant, a canonical phase space transformation is induced by such a transformation, \textcolor{black}{while} for the quantised system a transformation
in the unitary \textcolor{black}{configuration space} representation of the Heisenberg algebra being used for the canonically quantised system
\textcolor{black}{is induced as well}, in correlation with possible changes of local
quantum phases for configuration space wave function representations of abstract quantum states. Section~\ref{Sect3} then applies these different features
specifically when the system is coupled to a classical external background gauge field, whose gauge transformations induce transformations of the action
\textcolor{black}{precisely}
by additive total divergence terms. Especially for the canonically quantised system a careful assessment of the consequences of gauge transformations of the
classical background field on the quantised system must be addressed, to ascertain the gauge invariance and physical character of abstract quantum
observables and quantum states, irrespective of the choice of gauge for the background field. Section~\ref{Sect4} then considers \textcolor{black}{specifically}
the Landau problem,
first in its classical canonical formulation, as an illustrative example analysed in detail \textcolor{black}{in view of} the general considerations of the previous
two Sections.
And then by accounting as well for the role played by the global space-time symmetries of the Landau problem \textcolor{black}{and their conserved
Noether charges}. This analysis involves the most
general parametrisation possible for the choice of magnetic vector potential that must be specified for the local action of the Landau problem.
Finally Section~\ref{Sect5}, with its \textcolor{black}{five} Subsections \textcolor{black}{the last of which assesses the merits of Ref.\cite{Waka1}},
discusses at length the canonically quantised Landau problem, and what are the effects
of gauge transformations of the classical background vector potential for the quantised dynamics, paying due attention to the gauge invariant and physical
character of physical abstract quantum observables and quantum states, to be distinguished from those quantum operators that are not gauge invariant
and thus not physical. \textcolor{black}{Some} concluding comments are provided in Section~\ref{Sect6}.

\section{Canonical Quantisation and Choice of Lagrangian}
\label{Sect2}

Independently of the Landau problem but first now within a general context, let us consider a (mechanical) system whose configuration space manifold $M$
is spanned by local
curvilinear coordinates $q^n$ ($n=1,2,\cdots, N$), and whose dynamics derives through the variational principle from a local (in time) action in the form
of\footnote{For the purpose of the present discussion, only Lagrange functions that are not explicitly dependent on the dynamical evolution parameter $t$
are considered herein. Nevertheless the \textcolor{black}{discussion} hereafter \textcolor{black}{remains}
relevant even when the possibility of such a dependency on $t$ is included as well.
Even though here $t$ need not necessarily be the physical time as measured by some observer relative to a space-time reference frame,
this evolution parameter will be referred to as ``the time parameter''. In addition, it is implicitly assumed that the Lagrange functions considered herein
all define regular systems, with a regular Hessian. However all the same considerations of the present work readily extend to singular Lagrange functions
and their constrained dynamics \textcolor{black}{(see for example Ref.\cite{Gov1}}).},
\begin{equation}
S_1[q^n]=\int dt \,L_1(q^n(t),\dot{q}^n(t)).
\end{equation}

The canonical or Hamiltonian formalism that derives from the above Lagrangian formulation of such a dynamics includes, on the one hand,
a symplectic phase space spanned by canonical coordinates $(q^n,p^{(1)}_n)$ with the conjugate momenta $p^{(1)}_n$ defined by
\begin{equation}
p^{(1)}_n=\frac{\partial L_1}{\partial\dot{q}^n},
\end{equation}
whose Poisson brackets are canonical with the values
\begin{equation}
\left\{q^n,q^m\right\}=0,\qquad
\left\{q^n,p^{(1)}_m\right\}=\delta^n_m,\qquad
\left\{p^{(1)}_n,p^{(1)}_m\right\}=0,
\end{equation}
and on the other hand, the canonical Hamiltonian which generates time evolution in phase space through these Poisson brackets,
\begin{equation}
H_1(q^n,p^{(1)}_n)=\dot{q}^n p^{(1)}_n\,-\,L_1(q^n,\dot{q}^n)
\end{equation}
(under the present assumption of \textcolor{black}{the absence of any} explicit dependency in $t$ for $L_1(q^n,\dot{q}^n)$, the canonical Hamiltonian trivially
defines a conserved quantity, which is the Noether charge generator for constant translations in the time evolution parameter $t$).

However as is well known, given an ensemble of equations of motion a choice of \textcolor{black}{associated variational action and}
Lagrange function (when it exists) is determined only up to a total time derivative term
(or surface term in time). The same set of Euler-Lagrange equations of motion that derive from the above Lagrange function $L_1(q^n,\dot{q}^n)$ is also
obtained from the following alternative choice of action,
\begin{equation}S_2[q^n]=\int dt\,L_2(q^n,\dot{q}^n) \equiv \int dt\left(L_1(q^n,\dot{q}^n) + \frac{d}{dt}\Lambda(q^n)\right)
=\int dt\left(L_1(q^n,\dot{q}^n)+\dot{q}^n\frac{\partial}{\partial q^n}\Lambda(q^n)\right),
\end{equation}
where $\Lambda(q^n)$ is any \textcolor{black}{arbitrarily} chosen function defined over configuration space.
A new collection of phase space canonical coordinates $(q^n,p^{(2)}_n)$ ensues, with a different set of conjugate momenta such that
\begin{equation}
p^{(2)}_n=\frac{\partial L_2}{\partial\dot{q}^n}=\frac{\partial L_1}{\partial\dot{q}^n}+\frac{\partial}{\partial q^n}\Lambda(q^n)=p^{(1)}_n+\partial_n\Lambda(q^n),\qquad
p^{(1)}_n=p^{(2)}_2-\partial_n\Lambda(q^n),
\end{equation}
while the corresponding canonical Hamiltonian is obtained as,
\begin{equation}
H_2(q^n,p^{(2)}_n)=\dot{q}^n p^{(2)}_n - L_2(q^n,\dot{q}^n) = \dot{q}^np^{(1)}_n-L_1(q^n,\dot{q}^n)=H_1(q^n,p^{(1)}_n).
\end{equation}
Since furthermore the second phase space parametrisation $(q^n,p^{(2)}_n)$ is canonical as well with canonical Poisson brackets, namely,
\begin{equation}
\left\{q^n,q^m\right\}=0,\qquad
\left\{q^n,p^{(2)}_m\right\}=\delta^n_m,\qquad
\left\{p^{(2)}_n,p^{(2)}_m\right\}=0,
\end{equation}
these observations mean that any redefinition of some Lagrange function by an arbitrary total (time) derivative term in the form of
$L_2(q^n,\dot{q}^n)=L_1(q^n,\dot{q}^n)+\dot{q}^n\partial_n\Lambda(q^n)$, simply induces for the corresponding transformed
Hamiltonian formulation a canonical phase space transformation in the form of
\begin{equation}
q^n \rightarrow q^n,\quad
p^{(1)}_n \rightarrow p^{(2)}_n(q^n,p^{(1)}_n)=p^{(1)}_n + \partial_n\Lambda(q^n),\quad
H_1(q^n,p^{(1)}_n) \rightarrow H_2(q^n,p^{(2)}_n)=H_1(q^n,p^{(1)}_n),
\end{equation}
and this without affecting the \textcolor{black}{intrinsic, namely abstract classical} phase space dynamics.

A canonical quantisation of this dynamics derives from the correspondence principle. When based on the choice of action $S_1[q^n]$,
the Hilbert space of \textcolor{black}{abstract} quantum states provides a representation of the phase space Heisenberg algebra for the operators
$(\hat{q}^n,\hat{p}^{(1)}_n,\mathbb{I})$ (considered at some reference time $t=t_0$, which is not made explicit hereafter) in the form of,
\begin{equation}
\left[\hat{q}^n,\hat{q}^m\right]=0,\quad
\left[\hat{q}^n,\hat{p}^{(1)}_m\right]=i\hbar\,\delta^n_m\,\mathbb{I},\quad
\left[\hat{p}^{(1)}_n,\hat{p}^{(1)}_m\right]=0,\quad
\left(\hat{q}^n\right)^\dagger=\hat{q}^n,\quad
\left(\hat{p}^{(1)}_n\right)^\dagger=\hat{p}^{(1)}_n,
\end{equation}
while time evolution (whether in the Schr\"odinger picture in terms of the Schr\"odinger equation for \textcolor{black}{abstract} quantum states,
or in the Heisenberg picture in terms of the Heisenberg equation for \textcolor{black}{abstract} quantum observables) is governed by the quantum Hamiltonian
\begin{equation}
\hat{H}_1=H_1(\hat{q}^n,\hat{p}^{(1)}_n),
\end{equation}
given a choice of ordering for products of noncommuting operators contributing to composite quantum operators
(such that $\hat{H}_1$ be self-adjoint, or at least hermitian).

In configuration space, the general representation of such a Heisenberg algebra may be constructed as follows \cite{Gov1,Gov2}.
Since all the configuration space operators $\hat{q}^n$ commute with one another, consider a first \textcolor{black}{choice for some}
arbitrarily chosen complete basis of configuration
space eigenstates $|q^n;1\rangle$, such that $\hat{q}^n|q^n;1\rangle= q^n\,|q^n;1\rangle$, with their eigenvalues $q^n$ spanning the complete configuration space,
while \textcolor{black}{at this stage} their normalisation and local quantum phase \textcolor{black}{remain} free to be \textcolor{black}{adapted}
at one's best convenience. Let us fix their normalisation such that,
\begin{equation}
\langle q^n; 1 |{q'}^n;1 \rangle = \frac{1}{\sqrt{g(q^n)}}\,\delta^{(N)}(q^n-{q'}^n),\qquad
\mathbb{I}=\int_M d^N\!q^n\,\sqrt{g(q^n)}\,|q^n;1 \rangle \langle q^n;1 |,\qquad
g(q^n)>0,
\end{equation}
the quantity $g(q^n)>0$ being some specific but otherwise \textcolor{black}{some} {\sl a priori} arbitrary function over configuration space.
When configuration space is endowed with a Riemannian metric of tensor $g_{nm}(q^n)$, it is natural
to choose\footnote{In particular for an Euclidean configuration space \textcolor{black}{with $q^n$ being} cartesian coordinates,
one has $g(q^n)=1$.} $g(q^n)={\rm det}\,(g_{nm}(q^n))$,
thereby ensuring the covariance under configuration space diffeomorphisms of configuration space wave function representations of abstract
quantum states \textcolor{black}{$|\psi\rangle$, given by}
$\psi_{(1)}(q^n) \equiv \langle q^n;1|\psi\rangle$. Note well however, that any such choice of normalisation still leaves open the freedom to \textcolor{black}{adapt}
the local continuous quantum phase factor for the basis states $|q^n;1\rangle$\textcolor{black}{, thus through some local U(1) unitary transformation in configuration
space of this initially chosen eigenbasis of the abstract configuration space coordinate operators $\hat{q}^n$}.

Given any such choice of basis $|q^n;1\rangle$ and for whatever abstract quantum observable $\hat{\cal O}_1={\cal O}_1(\hat{q}^n,p^{(1)}_n)$, once the
matrix elements $\langle q^n;1|\hat{\cal O}_1|\psi\rangle$ for some abstract quantum state $|\psi\rangle$ are known, the action of that abstract
observable onto that abstract state is provided by the explicit representation
\begin{equation}
\hat{\cal O}_1|\psi\rangle = \int_M d^N\!q^n\sqrt{g(q^n)}\,|q^n;1\rangle\,\langle q^n;1|\hat{\cal O}_1|\psi\rangle.
\end{equation}

It may then be established \cite{Gov2} that relative to the basis $|q^n;1\rangle$, the most general \textcolor{black}{unitary configuration space}
representation of the Heisenberg algebra is provided by the following matrix elements, given any abstract quantum state $|\psi\rangle$
and its configuration space wave function \textcolor{black}{representation}
$\psi_{(1)}(q^n)\equiv \langle q^n;1|\psi\rangle$, 
\begin{equation}
\langle q^n;1 |\hat{q}^n |\psi\rangle = q^n\,\psi_{(1)}(q^n),\quad
\langle q^n;1 | \hat{p}^{(1)}_n | \psi\rangle = \frac{-i\hbar}{g^{1/4}(q^n)}\left(\frac{\partial}{\partial q^n} + \frac{i}{\hbar} V^{(1)}_n(q^n)\right)\,g^{1/4}(q^n)\,\psi_{(1)}(q^n).
\label{eq:Repr1}
\end{equation}
Here $V^{(1)}_n(q^n)$ is some arbitrarily chosen real vector field over configuration space which, however, needs to be of vanishing curl (or field strength), namely such that
\begin{equation}
V^{(1)}_{nm}(q^n) \equiv \partial_n V^{(1)}_m(q^n) - \partial_m V^{(1)}_n(q^n) = 0
\end{equation}
(this condition is required by the vanishing commutators of all the conjugate momenta $\hat{p}^{(1)}_n$ among themselves).

However given any \textcolor{black}{local} function $\chi_1(q^n)$ over configuration space, when transforming locally the still free to choose local quantum phase
of the configuration space eigenstates in the form of
\begin{equation}
|q^n;1,\chi_1\rangle \equiv U_{\chi_1}(\hat{q}^n)\,|q^n;1\rangle=e^{\frac{i}{\hbar}\chi_1(q^n)}|q^n;1\rangle,\qquad
\langle q^n;1,\chi_1| = \langle q^n;1|U^\dagger_{\chi_1}(\hat{q}^n)=e^{-\frac{i}{\hbar}\chi_1(q^n)}\langle q^n;1|,
\label{eq:Phase1}
\end{equation}
with the local U(1) unitary operator $U_{\chi_1}(\hat{q}^n)$ defined by
\begin{equation}
U_{\chi_1}(\hat{q}^n)\equiv e^{\frac{i}{\hbar}\chi_1(\hat{q}^n)},
\end{equation}
so that the corresponding configuration space wave function representations of \textcolor{black}{abstract quantum states} $|\psi\rangle$
transform as well in the form of\footnote{One could say that basis states $|q^n;1\rangle$ transform covariantly, whereas
\textcolor{black}{configuration space wave functions $\langle q^n;1|\psi\rangle$ of
abstract states $|\psi\rangle$ transform contravariantly, under the {\sl passive} unitary transformations $U_{\chi_1}(\hat{q}^n)$ acting on
the configuration space eigenbasis $|q^n; 1\rangle$.}}
\begin{equation}
\psi_{(1,\chi_1)}(q^n) \equiv \langle q^n;1,\chi_1 | \psi\rangle = \langle q^n;1|U^\dagger_{\chi_1}(\hat{q}^n)|\psi\rangle
 = e^{-\frac{i}{\hbar}\chi_1(q^n)}\,\psi_{(1)}(q^n),
\end{equation}
it readily follows that such a local U(1) redefinition of the local quantum phase factor of the basis of configuration space eigenstates induces indeed
a local U(1) gauge transformation of the vector field $V^{(1)}_n(q^n)$ which labels the choice of unitary \textcolor{black}{configuration
space} representation of the Heisenberg algebra,
with its configuration space matrix elements relative now to the \textcolor{black}{transformed} basis $|q^n;1,\chi_1\rangle$ then expressed as
\begin{eqnarray}
\langle q^n;1,\chi_1 |\hat{q}^n |\psi\rangle  &=& q^n\,\psi_{(1,\chi_1)}(q^n), \nonumber \\
\langle q^n;1,\chi_1 | \hat{p}^{(1)}_n | \psi\rangle  &=& \frac{-i\hbar}{g^{1/4}(q^n)}\left(\frac{\partial}{\partial q^n} + \frac{i}{\hbar} V^{(1,\chi_1)}_n(q^n)\right)\,g^{1/4}(q^n)\,\psi_{(1,\chi_1)}(q^n),
\label{eq:Repr3}
\end{eqnarray}
where $V^{(1,\chi_1)}_n(q^n)$ is given by,
\begin{equation}
V^{(1,\chi_1)}_n(q^n) = V^{(1)}_n(q^n) + \frac{\partial}{\partial q^n} \chi_1(q^n).
\end{equation}
In other words, unitarily inequivalent representations of the Heisenberg algebra are classified by the U(1) gauge equivalence classes of flat U(1) gauge connections
on the configuration space manifold, which are in one-to-one correspondence with the U(1) representations of the first homotopy group
\textcolor{black}{or fundamental group} of that manifold
(which are in turn labelled by the U(1) holonomies of $V^{(1)}_n(q^n)$ around the noncontractible cycles in the configuration manifold) \cite{Gov2}.

\textcolor{black}{In particular,} when that fundamental group is trivial, any flat U(1) connection is pure gauge, \textcolor{black}{namely}
$V^{(1)}_n(q^n)=\partial_n V^{(1)}(q^n)$ for some scalar field $V^{(1)}(q^n)$,
so that there always exists a choice of local quantum phase for the configuration space eigenstates such that $V^{(1,\chi_1)}_n(q^n)=0$
(with $\chi_1(q^n)=-V^{(1)}(q^n)$), thereby producing the canonical configuration space wave function representation of the Heisenberg algebra in the form of,
\begin{equation}
\langle q^n;1,\chi_1 |\hat{q}^n |\psi\rangle = q^n\,\psi_{(1,\chi_1)}(q^n),\ \ 
\langle q^n;1,\chi_1 | \hat{p}^{(1)}_n | \psi\rangle = \frac{-i\hbar}{g^{1/4}(q^n)} \frac{\partial}{\partial q^n}\,g^{1/4}(q^n)\,\psi_{(1,\chi_1)}(q^n).
\end{equation}
This observation corresponds to the (generalised) Stone-von Neumann theorem for unitary \textcolor{black}{configuration space}
representations of the Heisenberg algebra
(and thus applies, in particular, to Euclidean spaces, with then the unique well-known canonical
representation in cartesian coordinates, up to unitary phase transformations). Yet any arbitrary change in the local quantum phase of configuration space
eigenstates in the form of $|q^n;1,\chi_1\rangle\equiv U_{\chi_1}(\hat{q}^n)|q^n;1\rangle$\textcolor{black}{, namely a {\sl passive} U(1) unitary transformation of
the basis of configuration space eigenstates {\sl which leaves abstract quantum states and observables invariant},}
still goes hand-in-hand with an arbitrary choice of a nonvanishing flat U(1) gauge connection $V^{(1)}_n(q^n)$ and its own U(1) gauge transformation.
For instance \textcolor{black}{in particular, note well that} even
in the case of cartesian coordinates the canonical correspondence rule $\hat{p}^{(1)}_n\rightarrow -i\hbar\partial/\partial q^n$ assumes implicitly a specific choice
of local quantum phase for the configuration space eigenstates such that $V^{(1)}_n(q^n)=0$.

This fact needs to be kept in mind when using unitary \textcolor{black}{configuration space} representations of the Heisenberg
algebra obtained through different constructions for a same physical system and thus a same Hilbert space of \textcolor{black}{abstract} quantum states.
Different constructions may entail different implicit choices of local phase factors for the configuration space eigenstates,
thus different choices for the associated nonvanishing flat U(1) connection that provides a (wave) functional representation of the same abstract
conjugate momenta operators $\hat{p}^{(1)}_n$. \textcolor{black}{Generally, such a possibility} must be accounted for when moving
from one representation to another.
Even though all unitary \textcolor{black}{configuration space} representations are unitarily equivalent when the fundamental group of the configuration space is trivial,
abstract quantum states have a \textcolor{black}{configuration space} wave function representation relative not only to some choice of basis in Hilbert space---as is
the case for any abstract vector in some
Euclidean vector space over $\mathbb{R}$, even of finite dimension---, but in addition, the quantum phases involved in that representation are relative
as well to some choice of the quantum phase factors for that basis of quantum states in Hilbert space, as is indeed the case obviously in the present instance
for the configuration space wave function representations $\psi_{(1)}(q^n) \equiv \langle q^n;1|\psi\rangle$ and
$\psi_{(1,\chi_1)}(q^n) \equiv \langle q^n;1,\chi_1|\psi\rangle=\langle q^n;1|U^\dagger_{\chi_1}(\hat{q}^n)|\psi\rangle=e^{-i\chi_1(q^n)/\hbar}\,\psi_{(1)}(q^n)$
of any specific abstract quantum state $|\psi\rangle$.

When canonical quantisation proceeds from the Hamiltonian formulation based now on the action $S_2$ that differs from $S_1$ by a total time derivative surface term,
$S_2 - S_1=\int d\Lambda(q^n)$, exactly the same considerations apply {\it mutatis mutandis} to the Heisenberg algebra generated
by the \textcolor{black}{abstract quantum} operators $(\hat{q}^n,\hat{p}^{(2)}_n,\mathbb{I})$ with the corresponding \textcolor{black}{abstract} quantum Hamiltonian
\begin{equation}
\hat{H}_2=H_2(\hat{q}^n,\hat{p}^{(2)}_n)=H_1(\hat{q}^n,\hat{p}^{(1)}_n)=H_1(\hat{q}^n,\hat{p}^{(2)}_n-\partial_n\Lambda(\hat{q}^n))=\hat{H}_1,
\end{equation}
\textcolor{black}{which thus remains indeed invariant under the canonical transformation induced in phase space by the redefinition of the action by a surface term.}

An arbitrarily chosen complete basis of configuration space eigenstates
$|q^n;2\rangle$ may be normalised once again with the same choice of volume $N$-form factor $g(q^n)>0$ through
\begin{equation}
\langle q^n; 2 |{q'}^n;2 \rangle = \frac{1}{\sqrt{g(q^n)}}\,\delta^{(N)}(q^n-{q'}^n),\qquad
\mathbb{I}=\int_M d^N\!q^n\,\sqrt{g(q^n)}\,|q^n;2 \rangle \langle q^n;2 |,
\end{equation}
while their local quantum phase is still left to be \textcolor{black}{adapted} freely. The general configuration space wave function representation of the Heisenberg algebra
is parametrised by an arbitrary real vector field $V^{(2)}_n(q^n)$ of vanishing curl, $\partial_nV^{(2)}_m-\partial_m V^{(2)}_n=0$,
such that for any \textcolor{black}{abstract} quantum state $|\psi\rangle$ and its configuration space wave function \textcolor{black}{representation}
$\psi_{(2)}(q^n)\equiv \langle q^n;2|\psi\rangle$ one has,
\begin{equation}
\langle q^n;2 |\hat{q}^n |\psi\rangle = q^n\,\psi_{(2)}(q^n),\quad
\langle q^n;2 | \hat{p}^{(2)}_n | \psi\rangle = \frac{-i\hbar}{g^{1/4}(q^n)}\left(\frac{\partial}{\partial q^n} + \frac{i}{\hbar} V^{(2)}_n(q^n)\right)\,g^{1/4}(q^n)\,\psi_{(2)}(q^n).
\label{eq:Repr2}
\end{equation}
Under local quantum phase redefinitions of the configuration space eigenstates $|q^n;2\rangle$ involving an arbitrary function $\chi_2(q^n)$, such that
\begin{equation}
|q^n;2,\chi_2\rangle \equiv U_{\chi_2}(\hat{q}^n)\,|q^n;2\rangle=e^{\frac{i}{\hbar}\chi_2(q^n)}|q^n;2\rangle,\qquad
\langle q^n;2,\chi_2| = \langle q^n;2|U^\dagger_{\chi_2}(\hat{q}^n)=e^{-\frac{i}{\hbar}\chi_2(q^n)}\langle q^n;2|,
\label{eq:Phase2}
\end{equation}
so that the corresponding configuration space wave function representation of some abstract quantum state $|\psi\rangle$ transforms as well in the form of
\begin{equation}
\psi_{(2,\chi_2)}(q^n) \equiv \langle q^n;2,\chi_2 | \psi\rangle = \langle q^n;2|U^\dagger_{\chi_2}(\hat{q}^n)|\psi\rangle
 = e^{-\frac{i}{\hbar}\chi_2(q^n)}\,\psi_{(2)}(q^n),
\end{equation}
the vector field $V^{(2)}_n(q^n)$ transforms as a U(1) gauge connection over configuration space,
\begin{equation}
V^{(2,\chi_2)}_n(q^n) = V^{(2)}_n(q^n) + \frac{\partial}{\partial q^n} \chi_2(q^n).
\end{equation}
Hence given the \textcolor{black}{transformed} basis $|q^n; 2,\chi_2\rangle$ the Heisenberg algebra is \textcolor{black}{now represented} as,
\begin{eqnarray}
\langle q^n;2,\chi_2 |\hat{q}^n |\psi\rangle &=&  q^n\,\psi_{(2,\chi_2)}(q^n), \nonumber \\
\langle q^n;2,\chi_2 | \hat{p}^{(2)}_n | \psi\rangle &=& \frac{-i\hbar}{g^{1/4}(q^n)}\left(\frac{\partial}{\partial q^n} + \frac{i}{\hbar} V^{(2,\chi_2)}_n(q^n)\right)\,g^{1/4}(q^n)\,\psi_{(2,\chi_2)}(q^n).
\label{eq:Repr4}
\end{eqnarray}

The two actions $S_1$ and $S_2$ determine the dynamics of a same physical system, for which the abstract Hilbert space providing a representation
of their respective Heisenberg algebras must be identical, with in particular the same \textcolor{black}{abstract} configuration space operators $\hat{q}^n$.
Yet, the total time derivative contribution by which these two actions differ implies in fact a canonical transformation leading to distinct conjugate momenta
in classical phase such that $p^{(2)}_n=p^{(1)}_n+\partial_n\Lambda(q^n)$, but identical \textcolor{black}{abstract} Hamiltonians nonetheless,
$H_2(q^n,p^{(2)}_n)=H_1(q^n,p^{(1)}_n)=H_1(q^n,p^{(2)}_n-\partial_n\Lambda(q^n))$.

For the quantised system the \textcolor{black}{abstract} conjugate momenta operators are thus related as
\begin{equation}
\hat{p}^{(2)}_n=\hat{p}^{(1)}_n+\partial_n\Lambda(\hat{q}^n)=e^{-\frac{i}{\hbar}\Lambda(\hat{q}^n)}\,\hat{p}^{(1)}_n\,e^{\frac{i}{\hbar}\Lambda(\hat{q}^n)}
=U^\dagger_\Lambda(\hat{q}^n)\,\hat{p}^{(1)}_n\,U_\Lambda(\hat{q}^n),\quad
\hat{p}^{(1)}_n=U_\Lambda(\hat{q}^n)\,\hat{p}^{(2)}_n\,U^\dagger_\Lambda(\hat{q}^n),
\end{equation}
where $U_\Lambda(\hat{q}^n)=e^{\frac{i}{\hbar}\Lambda(\hat{q}^n)}$ is the unitary operator generated by the change of action through
the surface term $\int d\Lambda$, which is also such that
\begin{equation}
U^\dagger_\Lambda(\hat{q}^n)\,\hat{q}^n\,U_\Lambda(\hat{q}^n) = \hat{q}^n,
\end{equation}
thus leaving invariant the \textcolor{black}{abstract} configuration space operators for the two canonical quantisations.
\textcolor{black}{Note well however that even though not invariant under the canonical transformation generated by the surface term in $\Lambda$,
the abstract quantum canonical conjugate momenta $\hat{p}^{(1)}_n$ and $\hat{p}_n^{(2)}$ transform covariantly under $U_\Lambda(\hat{q}^n)$.}

In contradistinction, \textcolor{black}{one has} for the quantum Hamiltonians,
\begin{equation}
\hat{H}_2 = H_2(\hat{q}^n,\hat{p}^{(2)}_n)=H_1(\hat{q}^n,\hat{p}^{(1)}_n) = \hat{H}_1,
\end{equation}
\textcolor{black}{thus defining a same and unique abstract quantum observable, $\hat{H}$.}
Nevertheless the following relation applies between \textcolor{black}{the different functional realisations of this same abstract operator},
\begin{equation}
\hat{H}=\hat{H}_2=\hat{H}_1=H_1(\hat{q}^n,\hat{p}^{(1)}_n)=H_1(\hat{q}^n,U_\Lambda(\hat{q}^n) \hat{p}^{(2)}_n U^\dagger_\Lambda(\hat{q}^n))
=U_\Lambda(\hat{q}^n) H_1(\hat{q}^n,\hat{p}^{(2)}_n) U^\dagger_\Lambda(\hat{q}^n).
\end{equation}
This same observation extends to any \textcolor{black}{abstract}
quantum observable \textcolor{black}{such that} $\hat{\cal O}_2={\cal O}_2(\hat{q}^n,\hat{p}^{(2)}_n)={\cal O}_1(\hat{q}^n,\hat{p}^{(1)}_n)=\hat{\cal O}_1$,
with again the following likewise relation between these \textcolor{black}{different functional realisations of the operator, thus indeed defining
a same and unique abstract quantum observable,}
\begin{equation}
\hat{\cal O}=\hat{\cal O}_2=\hat{\cal O}_1={\cal O}_1(\hat{q}^n,\hat{p}^{(1)}_n)
={\cal O}_1(\hat{q}^n,U_\Lambda(\hat{q}^n) \hat{p}^{(2)}_n U^\dagger_\Lambda(\hat{q}^n))
=U_\Lambda(\hat{q}^n)\,{\cal O}_1(\hat{q}^n,\hat{p}^{(2)}_n)\,U^\dagger_\Lambda(\hat{q}^n).
\label{eq:Obser}
\end{equation}

Since the two canonical quantisations share the same \textcolor{black}{abstract} configuration space operators $\hat{q}^n$,
let us consider more explicitly the possible local quantum phase
difference between the two choices of configuration space eigenstates $|q^n;1\rangle$ and $|q^n;2\rangle$, in the form of
\begin{equation}
|q^n;2\rangle = e^{\frac{i}{\hbar}\chi_{21}(q^n)}\,|q^n;1\rangle=U_{\chi_{21}}(\hat{q}^n)|q^n;1\rangle,\quad
\langle q^n;2| = e^{-\frac{i}{\hbar}\chi_{21}(q^n)}\,\langle q^n;1|=\langle q^n;1| U^\dagger_{\chi_{21}}(\hat{q}^n),
\end{equation}
where $\chi_{21}(q^n)$ is thus some specific function over configuration space such that
\begin{equation}
\langle q^n;1|{q'}^n;2\rangle=\frac{1}{\sqrt{g(q^n)}}\,e^{\frac{i}{\hbar}\chi_{21}(q^n)}\,\delta^{(N)}(q^n-{q'}^n).
\end{equation}
Consequently the configuration space wave function representations of a same abstract
quantum state $|\psi\rangle$ for each of the two canonical quantisations based on $S_1$ and $S_1$ are related as,
\begin{equation}
\psi_{(2)}(q^n) = \langle q^n;2|\psi\rangle = \langle q^n;1|U^\dagger_{\chi_{21}}(\hat{q}^n)|\psi\rangle
= \langle q^n;1|\psi;\chi_{21}\rangle =e^{-\frac{i}{\hbar}\chi_{21}(q^n)}\,\psi_{(1)}(q^n).
\end{equation}
Quite obviously the correspondence rule for the two configuration space wave function representations of the operators $\hat{q}^n$ remains identical, with
\begin{equation}
\langle q^n;2|\hat{q}^n|\psi\rangle = q^n\,\psi_{(2)}(q^n),\qquad
\langle q^n;1|\hat{q}^n|\psi\rangle = q^n\,\psi_{(1)}(q^n).
\end{equation}
However, the situation for the conjugate momenta $\hat{p}^{(1)}_n$ and $\hat{p}^{(2)}_n$ is necessarily different,
since $\hat{p}^{(2)}_n-\hat{p}^{(1)}_n=\partial_n\Lambda(\hat{q}^n)$. One finds that the two flat U(1) connections $V^{(1)}_n(q^n)$ and $V^{(2)}_n(q^n)$
characterising each of the two Heisenberg algebra unitary representations for the conjugate momenta, in the form of (\ref{eq:Repr1},\ref{eq:Repr2}), are
related by the following U(1) gauge transformation which combines a transformation generated by the action surface term in $\Lambda(q^n)$ with a transformation
generated by the local relative quantum phase difference $\chi_{21}(q^n)$,
\begin{equation}
V^{(2)}_n(q^n) = V^{(1)}_n(q^n) + \partial_n \Lambda(q^n) + \partial_n \chi_{21}(q^n).
\end{equation}
More generally if in addition the local quantum phases of the basis states $|q^n;1\rangle$ and $|q^n;2\rangle$ are modified as well
according to the transformations given in (\ref{eq:Phase1},\ref{eq:Phase2}), the transformed flat U(1) connections $V^{(1,\chi_1)}_n(q^n)$
and $V^{(2,\chi_2)}_n(q^n)$ are then related as,
\begin{equation}
V^{(2,\chi_2)}_n(q^n)=V^{(1,\chi_1)}_n(q^n)+\partial_n\Lambda(q^n)+\partial_n\left(\chi_{21}(q^n) + \chi_2(q^n) - \chi_1(q^n)\right),
\label{eq:generalV}
\end{equation}
with the respective representations of the conjugate momenta operators $\hat{p}^{(1)}_n$ and $\hat{p}^{(2)}_n$ given in (\ref{eq:Repr3},\ref{eq:Repr4}).

In other words, while at the classical level a redefinition of the action by a time surface term $\int d\Lambda$ only induces a canonical transformation
of the conjugate momenta with $p^{(2)}_n=p^{(1)}_n+\partial_n\Lambda(q^n)$, for the canonically quantised system such a redefinition induces not only
the same transformation of the \textcolor{black}{abstract} conjugate momenta operators which is effected through the \textcolor{black}{covariant action of the}
U(1) unitary operator $U_\Lambda(\hat{q}^n)$, but
in addition it also induces a unitary change in the unitary representation of the Heisenberg algebra that is being used, effectively through a specific U(1) gauge
transformation of the flat U(1) connection that parametrises such representations, this gauge transformation being determined as well
directly by the surface term $\Lambda$.
Note well however, that this gauge transformation induced by $\Lambda$ does not change the U(1) gauge equivalence class of the Heisenberg algebra unitary
representation that is being used (namely, the corresponding U(1) representation of the first homotopy group of the configuration space manifold).
The \textcolor{black}{abstract} quantum Hamiltonians being identical, $\hat{H}_2=\hat{H}_1$, and more generally any \textcolor{black}{abstract}
quantum observables, $\hat{\cal O}_2=\hat{\cal O}_1$,
this observation implies that the quantum dynamics of the system, as is readily the case of the classical system, remains itself independent as well
of the choice of surface term $\int d\Lambda$ for its action.

Yet it must \textcolor{black}{certainly} be kept in mind that by including such a surface term in time, in effect a unitary \textcolor{black}{configuration space}
representation of the Heisenberg algebra is being selected which is
different---with a different flat U(1) connection to represent the \textcolor{black}{abstract canonical} conjugate momenta operators---from that which is selected
in the absence of that surface term,
even though these two representations are unitarily equivalent. To be explicit, consider any arbitrary \textcolor{black}{abstract}
quantum observable $\hat{\cal O}=\hat{\cal O}_1=\hat{\cal O}_2$
and its matrix element for any two arbitrary abstract quantum states $|\psi\rangle$ and $|\varphi\rangle$. Given that
\begin{equation}
\hat{\cal O}=\hat{\cal O}_2=\hat{\cal O}_1={\cal O}_1(\hat{q}^n,\hat{p}^{(1)}_n)
=U_\Lambda(\hat{q}^n)\,{\cal O}_1(\hat{q}^n,\hat{p}^{(2)}_n)\,U^\dagger_\Lambda(\hat{q}^n),
\end{equation}
evaluating the matrix element $\langle\psi | \hat{\cal O}|\varphi\rangle$ whether in the arbitrary $|q^n;1,\chi_1\rangle$ or $|q^n;2,\chi_2\rangle$ bases,
one has, respectively,
\begin{eqnarray}
\langle\psi | \hat{\cal O}|\varphi\rangle &=& \langle\psi|{\cal O}_1(\hat{q}^n,\hat{p}^{(1)}_n)|\varphi\rangle \\
&=& \int_M d^N\!q^n\,\psi^*_{(1,\chi_1)}(q^n)\,{\cal O}_1\left(q^n,\frac{-i\hbar}{g^{1/4}(q^n)}\left(\partial_n+\frac{i}{\hbar}V^{(1,\chi_1)}_n(q^n)\right)g^{1/4}(q^n)\right)\,
\varphi_{(1,\chi_1)}(q^n), \nonumber
\end{eqnarray}
as well as,
\begin{eqnarray}
\langle\psi | \hat{\cal O}|\varphi\rangle &=& \langle\psi|U_{\Lambda}(\hat{q}^n){\cal O}_1(\hat{q}^n,\hat{p}^{(2)}_n) U^\dagger_{\Lambda}(\hat{q}^n)|\varphi\rangle \\
&=& \int_M d^N\!q^n\,\psi^*_{(2,\chi_2)}(q^n) \nonumber \\
&&\quad \times
\,e^{\frac{i}{\hbar}\Lambda(q^n)}\,
{\cal O}_1\left(q^n,\frac{-i\hbar}{g^{1/4}(q^n)}\left(\partial_n+\frac{i}{\hbar}V^{(2,\chi_2)}_n(q^n)\right) g^{1/4}(q^n)\right)\,e^{-\frac{i}{\hbar}\Lambda(q^n)}\,
\varphi_{(2,\chi_2)}(q^n), \nonumber
\end{eqnarray}
with the flat U(1) gauge connections characterising the two unitary representations of the Heisenberg algebras involved, which are related through
the U(1) gauge transformation determined in~(\ref{eq:generalV}),
\begin{equation}
V^{(2,\chi_2)}_n(q^n)=V^{(1,\chi_1)}_n(q^n)+\partial_n\left(\Lambda(q^n)+\chi_{21}(q^n) + \chi_2(q^n) - \chi_1(q^n)\right).
\end{equation}
However, since furthermore one has the following local quantum phase relations for the relevant wave function representations of the abstract quantum states,
\begin{eqnarray}
\psi_{(2,\chi_2)}(q^n) &=& e^{-\frac{i}{\hbar}(\chi_{21}(q^n)+\chi_2(q^n)-\chi_1(q^n))}\,\psi_{(1,\chi_1)}(q^n), \nonumber \\
\varphi_{(2,\chi_2)}(q^n) &=& e^{-\frac{i}{\hbar}(\chi_{21}(q^n)+\chi_2(q^n)-\chi_1(q^n))}\,\varphi_{(1,\chi_1)}(q^n),
\end{eqnarray}
the above two integral representations of the same abstract matrix element $\langle\psi|\hat{\cal O}|\varphi\rangle$ are indeed of exactly identical values.

In other words, in order to ascertain that the same quantum dynamics is still accounted for when transforming its action principle by a total time
derivative surface term, $\int d\Lambda$,
due care must be exercised in properly keeping track of the induced unitary transformation between the unitary \textcolor{black}{configuration space}
representations of the Heisenberg
algebra that are being used. This must be done concurrently with the proper consideration for the values of the local quantum phases for the bases of configuration space eigenstates that are being chosen
to represent abstract quantum states in terms of configuration space wave functions. In particular, the U(1) gauge transformations thereby induced
all belong to the same U(1) gauge equivalence class of the identity transformation, since all scalar fields $\Lambda(q^n)$, $\chi_{21}(q^n)$, $\chi_2(q^n)$ and
$\chi_1(q^n)$ are single-valued over configuration space. Hence, the flat U(1) gauge connections involved, \textcolor{black}{namely} $V^{(1,\chi_1)}_n(q^n)$ and
$V^{(2,\chi_2)}_n(q^n)$, which parametrise distinct unitary \textcolor{black}{configuration space} representations
of the Heisenberg algebra and in particular of their \textcolor{black}{abstract canonical} conjugate momenta operators, all belong
to the same U(1) gauge equivalence class of unitarily inequivalent representations of the Heisenberg algebra
(in the case of a configuration space manifold of trivial fundamental group, there exists only a single such class of unitarily inequivalent representations).

Incidentally, note well that the above most general observation regarding the wave function representations of the matrix element
$\langle\psi|\hat{\cal O}|\varphi\rangle$ and their invariance under the U(1) gauge transformations generated by the surface term in $\Lambda(q^n)$,
also means that when these more subtle aspects are properly accounted for, there is certainly no need whatsoever to check, for any given observable,
that all its matrix elements for all the elements of whatever choice of basis in Hilbert space maintain an identical value whether the system's action is
transformed or not by a total time derivative surface term $\int d\Lambda$.

For instance, even when the fundamental group of the configuration space manifold is
trivial so that, (1) all \textcolor{black}{configuration space}
representations of the Heisenberg algebra are unitarily equivalent and one may choose simply a vanishing flat U(1) connection, say
in the absence of the surface term, namely $V^{(1)}_n(q^n)=0$ thus corresponding to the canonical choice of Heisenberg algebra representation;
while furthermore, (2) exactly the same bases of configuration space eigenstates
are chosen for each of the canonical quantisations, namely $|q^n;2\rangle=|q^n;1\rangle=|q^n\rangle$ so that $\chi_{21}(q^n)=0$
and $\psi_{(2)}(q^n)=\psi_{(1)}(q^n)=\psi(q^n)$; and while, (3) no other possible local quantum phase transformations are effected,
\textcolor{black}{namely} $\chi_1(q^n)=0=\chi_2(q^n)$;
(4) it remains so that the flat and pure gauge U(1) connection in presence of the surface term $\Lambda(q^n)$ certainly does not vanish,
but rather is given by,
\begin{equation}
V^{(2)}_n(q^n)=\partial_n\Lambda(q^n).
\end{equation}
If one were to overlook this fact and choose to represent the \textcolor{black}{abstract canonical} conjugate momenta operators in presence of the surface term
still with the canonical representation such that $V^{(2)}_n(q^n)=0$ (and the same wave function $\psi(q^n)$),
then clearly the corresponding two quantum formulations would be physically distinct,
leading to different matrix elements for the same physical observables whose wave function representations involve differentials in configuration space.
Alternatively one could combine the introduction of the surface term in $\Lambda(q^n)$ with changes in the local quantum phases of the configuration space
eigenstates such that
\begin{equation}
\partial_n\left(\Lambda(q^n) + \chi_{21}(q^n) + \chi_2(q^n) - \chi_1(q^n)\right) = 0,
\end{equation}
so that if the canonical choice is made for $V^{(1,\chi_1)}_n(q^n)=0$ one still has the canonical Heisenberg algebra representation with $V^{(2,\chi_2)}_n(q^n)=0$
for the quantum system in presence of the surface term $\Lambda(q^n)$.
However, in order then to meet the above condition one must take proper account of the appropriate local quantum phase redefinitions
of the configuration space wave functions in use, for example with $\chi_{21}(q^n)=-\Lambda(q^n)$ while $\chi_1(q^n)=0=\chi_2(q^n)$,
namely by choosing the basis $|q^n;2\rangle=e^{-\frac{i}{\hbar}\Lambda(q^n)}|q^n;1\rangle$ in combination with the change of action defined by the surface
term $\Lambda(q^n)$.
Otherwise once again matrix elements of operators whose wave function representations involve differentials in configuration space
would take physically distinct values when the surface term in $\Lambda$ is introduced.

\textcolor{black}{Hence even though the system's degrees of freedom are then left strictly invariant,}
changing an action by a total divergence (in time, and more generally in space-time in the case of field theories) certainly has subtle and non trivial
consequences for the quantised system and specifically for the unitary \textcolor{black}{configuration space}
representations to be chosen for its Heisenberg algebra,
that must properly be accounted for to ascertain a strictly physically identical quantum dynamics.

Using developments available in Ref.\cite{Gov2} involving in addition the eigenstates of the \textcolor{black}{abstract}
conjugate momenta operators $\hat{p}^{(1)}_n$ and $\hat{p}^{(2)}_n$,
further relevant considerations to complement those presented above regarding these different issues could be approached as well from the perspective
of a path integral or functional integral representation of quantum amplitudes, involving whether the action $S_1$ or the action $S_2$.
However such considerations are not included herein, since they are not directly relevant to the remainder of this work, in particular
\textcolor{black}{in the case of} the Landau problem.

To conclude, it is fair to comment that usually when canonically quantising an action principle, one is content with keeping to that specific action,
without transforming it by a local (time or space-time) surface term, in which case the different issues addressed above do not arise.
However this is no longer the case when coupling the dynamics to background fields which themselves are defined up to local gauge transformations,
that in turn induce a transformation of the action precisely by such a local surface or divergence term \textcolor{black}{while degrees of freedom
are left strictly invariant}. This is the case for instance with the Landau problem.
Consequences of such couplings still \textcolor{black}{first} in a general context \textcolor{black}{are addressed in} the next Section.

\section{Background Gauge Fields and Choice of Gauge}
\label{Sect3}

Let us first consider again some general (mechanical) system with a configuration space manifold $M$ spanned by local curvilinear coordinates $q^n$ ($n=1,2,\cdots,N$), whose
dynamics derives from a local Lagrangian action principle in the form of
\begin{equation}
S_0[q^n]=\int dt\,L_0(q^n,\dot{q}^n).
\end{equation}
Its Hamiltonian first-order action then reads,
\begin{equation}
S_0[q^n,p_n]=\int dt\left(\dot{q}^n p_n - H_0(q^n,p_n)\right),
\end{equation}
whose symplectic phase space is spanned by canonical coordinates $(q^n,p_n)$ with conjugate momenta defined as
\begin{equation}
p_n=\frac{\partial L_0}{\partial\dot{q}^n},
\end{equation}
and a canonical Hamiltonian given as,
\begin{equation}
H_0(q^n,p_n)=\dot{q}^n p_n - L_0(q^n,\dot{q}^n).
\end{equation}
Within this first-order formalism, time evolution is represented by the first-order Hamiltonian equations of motion,
\begin{equation}
\dot{q}^n=\frac{\partial H_0}{\partial p_n},\qquad
\dot{p}_n=-\frac{\partial H_0}{\partial q^n}.
\end{equation}

Let us now couple this dynamics whether to a real vector field $A^{(1)}_n(q^n)\textcolor{black}{=A^{(\alpha=1)}_n(q^n)}$,
or to another real vector field $A^{(2)}_n(q^n)\textcolor{black}{=A^{(\alpha=2)}_n(q^n)}$,
in either case through an action principle given as,
\begin{equation}
S_\alpha[q^n;A^{(\alpha)}_n]=\int dt \, L_\alpha(q^n,\dot{q}^n;A^{(\alpha)}_n),\quad
L_\alpha(q^n,\dot{q}^n;A^{(\alpha)}_n)=L_0(q^n,\dot{q}^n) + \dot{q}^n A^{(\alpha)}_n(q^n),\quad \alpha=1,2.
\end{equation}
Correspondingly one has two field strength tensors denoted
\begin{equation}
A^{(\alpha)}_{nm}(q^n)=\partial_n A^{(\alpha)}_m(q^n) - \partial_m A^{(\alpha)}_n(q^n),\qquad \alpha=1,2.
\end{equation}

The Hamiltonian formulations of these two systems involve canonical phase space coordinates $(q^n,\pi[A^{(\alpha)}_n]_n)$
of which the conjugate momenta \textcolor{black}{$\pi[A^{(\alpha)}_n]_n$}, related to the momenta $p_n$ of the decoupled system through a simple shift
in the relevant vector field, are given as
\begin{equation}
\pi[A^{(\alpha)}_n]_n=\frac{\partial L_\alpha}{\partial\dot{q}^n}=p_n+A^{(\alpha)}_n(q^n),\qquad
p_n=\pi[A^{(\alpha)}_n]_n - A^{(\alpha)}_n(q^n),\qquad
\alpha=1,2.
\end{equation}
Being canonical these phase space coordinates possess the following Poisson brackets,
\begin{equation}
\left\{q^n,q^m\right\}=0,\qquad
\left\{q^n,\pi[A^{(\alpha)}_n]_m\right\}=\delta^n_m,\qquad
\left\{\pi[A^{(\alpha)}_n]_n,\pi[A^{(\alpha)}_n]_m\right\}=0,\qquad \alpha=1,2,
\end{equation}
hence also with the following values for the momenta $p_n$,
\begin{equation}
\left\{q^n,q^m\right\}=0,\qquad
\left\{q^n,p_m\right\}=\delta^n_m,\qquad
\left\{p_n,p_m\right\}=A^{(\alpha)}_{nm}(q^n),\qquad \alpha=1,2.
\end{equation}
Note well how the momenta $p_n$ now have nonvanishing Poisson brackets whose values coincide with the field strength $A^{(\alpha)}_{nm}(q^n)$
for any one choice of the two systems.

Furthermore the canonical Hamiltonians of the two background field coupled systems are obtained in the form of, respectively,
\begin{equation}
H_\alpha\left(q^n,\pi[A^{(\alpha)}_n]\right)=\dot{q}^n \pi[A^{(\alpha)}_n]_n - L_\alpha = H_0(q^n,p_n)=H_0\left(q^n,\pi[A^{(\alpha)}_n]_n-A^{(\alpha)}_n(q^n)\right),
\quad \alpha=1,2,
\end{equation}
thereby implying the following first-order Hamiltonian equations of motion in each case,
\begin{equation}
\dot{q}^n=\frac{\partial H_0}{\partial p_n},\qquad
\dot{p}_n=A^{(\alpha)}_{nm}(q^n)\,\frac{\partial H_0}{\partial p_m} - \frac{\partial H_0}{\partial q^n},\qquad
\alpha=1,2.
\end{equation}
Hence if $A^{(\alpha)}_{nm}(q^n)\ne 0$ the presence of the background vector field implies a dynamics different from what it is in the absence of these fields,
while if $A^{(2)}_{nm}(q^n)\ne A^{(1)}_{nm}(q^n)$ that dynamics is physically distinct for the two background field coupled systems with $\alpha=1$ and $\alpha=2$.

In contradistinction, if the two vector fields are related through a simple local gauge transformation of the form
\begin{equation}
A^{(2)}_n(q^n)= A^{(1)}_n(q^n) + \frac{\partial}{\partial q^n}\varphi(q^n),
\label{eq:gauge}
\end{equation}
with $\varphi(q^n)$ some scalar function over configuration space, even though different the two background vector gauge fields possess the same
field strength, $A^{(2)}_{nm}(q^n)=A^{(1)}_{nm}(q^n)=A_{nm}(q^n)$, and the two classical dynamics are then physically identical.
Since under such circumstance the Lagrange functions of the two actions $S_\alpha[q^n;A^{(\alpha)}_n]$ differ only by a total time derivative surface term
$\int d\varphi$, \textcolor{black}{with for these Lagrange functions,}
\begin{equation}
L_2(q^n,\dot{q}^n)=L_1(q^n,\dot{q}^n)+\dot{q}^n\partial_n\varphi(q^n)=L_1(q^n,\dot{q}^n)+\frac{d}{dt}\varphi(q^n),
\end{equation}
this observation implies \textcolor{black}{that one may at once} call on all the considerations presented in Section~\ref{Sect2}, inclusive of the same notations.

Thus the \textcolor{black}{canonical} conjugate momenta associated to any one of the two actions $S_\alpha[q^n;A^{(\alpha)}_n]$ accounting for the same physics differ
by the \textcolor{black}{contribution in $\partial_n\varphi$ resulting from the gauge transformation characterised by the scalar function $\varphi(q^n)$},
\begin{equation}
\pi[A^{(2)}_n]_n=\pi[A^{(1)}_n]_n + \partial_n\varphi(q^n),\qquad
\pi[A^{(1)}_n]_n=\pi[A^{(2)}_n]_n-\partial_n \varphi(q^n),
\end{equation}
while the momenta $p_n$ are given by the difference of each ensemble of these \textcolor{black}{canonical}
conjugate momenta with the corresponding background vector gauge field,
\begin{equation}
p_n=\pi[A^{(2)}_n]_n-A^{(2)}_n(q^n)=\pi[A^{(1)}_n]_n-A^{(1)}_n(q^n),
\end{equation}
whose Poisson brackets thus read,
\begin{equation}
\left\{q^n,q^m\right\}=0,\qquad
\left\{q^n,p_m\right\}=\delta^n_m,\qquad
\left\{p_n,p_m\right\}=A_{nm}(q^n).
\end{equation}
\textcolor{black}{In particular note well that in contradistinction to the canonical conjugate moment $\pi[A^{(\alpha)}_n]_n$ which are gauge variant,
the momenta $p_n$ are strictly invariant under gauge transformations of the background vector field.}

Furthermore the two \textcolor{black}{abstract} Hamiltonians also coincide through these changes of phase space variables
\textcolor{black}{in their functional phase space realisations}, and are given by $H_0(q^n,p_n)$ even in presence of the background fields,
\begin{equation}
H_2\left(q^n,\pi[A^{(2)}_n]_n\right)=H_0(q^n,p_n)=H_0(q^n,\pi[A^{(\alpha)}_n]_n-A^{(\alpha)}_n(q^n))=H_1\left(q^n,\pi[A^{(1)}_n]_n\right).
\end{equation}

It is worthwhile to emphasize the distinctive nature of such a gauge invariance, which must truly be distinguished from a symmetry in the ordinary sense,
namely a symmetry transformation {\it which acts on the degrees of freedom of a system} such that its equations of motion transform covariantly,
or are left invariant altogether.
Indeed note well that {\it the gauge transformation (\ref{eq:gauge}) acts only on that background vector gauge field},
but that the configuration space degrees of freedom of the system $q^n$ are left \textcolor{black}{strictly} invariant under that transformation. Nevertheless such a gauge
transformation of the background vector gauge field, by inducing a variation of the action through a time surface term, $\int d\varphi$, induces concurrently
a canonical transformation in its symplectic phase space that shifts the canonical \textcolor{black}{conjugate} momenta by the gauge transformation.
Consequently even though
the degrees of freedom $q^n$ are gauge invariant under gauge transformations of the background vector gauge field, their canonically conjugate momenta
$\pi[A^{(\alpha)}_n]_n$ are gauge variant and are dependent on that field. Yet by subtracting from these \textcolor{black}{canonical}
conjugate momenta the associated background
vector gauge field $A^{(\alpha)}_n(q^n)$, one obtains the momenta $p_n$ which are again
gauge invariant observables, thus also independent of the background vector gauge field, namely of the choice of gauge for the vector field accounting
for the local coupling of the system to the background field strength $A_{nm}(q^n)=A^{(\alpha)}_{nm}(q^n)$. The phase space parametrisation
using $(q^n,p_n)$ is thus invariant under gauge transformations of the background vector gauge field, and independent from the choice of gauge
for $A^{(\alpha)}_n(q^n)$. Consequently the Hamiltonian of the system, $H=H_\alpha(q^n,\pi[A^{(\alpha)}_n]_n)=H_0(q^n,p_n)$, is also a gauge invariant
observable in that specific sense, namely under gauge transformations \textcolor{black}{solely} of the \textcolor{black}{external}
background vector field (\textcolor{black}{which do not act on}
the system's degrees of freedom $q^n$). More generally \textcolor{black}{for any observable ${\cal O}_0(q^n,p_n)$ of the original system with action $S_0[q^n]$,}
any such gauge invariant observable ${\cal O}$ is given in the form of
\begin{equation}
{\cal O}={\cal O}_\alpha(q^n,\pi[A^{(\alpha)}_n]_n)={\cal O}_0(q^n,p_n)={\cal O}_0(q^n,\pi[A^{(\alpha)}_n]_n-A^{(\alpha)}_n(q^n)),\qquad \alpha=1,2.
\end{equation}

In contradistinction, any dependency on the conjugate momenta $\pi[A^{(\alpha)}_n]_n$ which is not \textcolor{black}{specifically}
through the combination $\pi[A^{(\alpha)}_n]_n-A^{(\alpha)}_n(q^n)=p_n$ is simply not gauge invariant in that sense,
and thus cannot be deemed to represent a physical and gauge invariant observable. In particular it would certainly be function of the choice of gauge
for the background vector field, as are the $\pi[A^{(\alpha)}_n]_n$'s themselves to begin with.
Finally in addition to the above consequences of gauge transformations in the background vector gauge field  for the classical dynamics, for the canonically
quantised system such gauge transformations thus induce as well transformations in the unitary \textcolor{black}{configuration space}
representation of the Heisenberg algebra to be used for \textcolor{black}{abstract}
quantum operators, and this in direct correlation with transformations in the choices of local quantum phases of the configuration space eigenstates to be used
for configuration space wave function representations of abstract quantum states $|\psi\rangle$, in order to account for a quantum dynamics which
remains independent as well from the choice of gauge for the background vector gauge field $A^{(\alpha)}_n(q^n)$, as discussed in Section~\ref{Sect2}.

Turning now to the canonically quantised system with its Heisenberg algebra of commutation relations (specified at some implicit reference time $t=t_0$),
one has for the \textcolor{black}{abstract quantum canonical phase space coordinate operators} $(\hat{q}^n,\hat{\pi}[A^{(\alpha)}_n]_n,\mathbb{I})$,
\begin{equation}
\left[\hat{q}^n,\hat{q}^m\right]=0,\qquad
\left[\hat{q}^n,\hat{\pi}[A^{(\alpha)}_n]_m\right]=i\hbar\,\delta^n_m\,\mathbb{I},\qquad
\left[\hat{\pi}[A^{(\alpha)}_n]_n,\hat{\pi}[A^{(\alpha)}_n]_m\right]=0,
\end{equation}
or for the \textcolor{black}{abstract} gauge invariant quantum operators $(\hat{q}^n,\hat{p}_n,\mathbb{I})$,
\begin{equation}
\left[\hat{q}^n,\hat{q}^m\right]=0,\qquad
\left[\hat{q}^n,\hat{p}_m\right]=i\hbar\,\delta^n_m\,\mathbb{I},\qquad
\left[\hat{p}_n,\hat{p}_m\right]=i\hbar\,A_{nm}(\hat{q}^n),
\end{equation}
together with the required hermiticity properties,
\begin{equation}
\hat{q}^{n^\dagger}=\hat{q}^n,\qquad
\hat{\pi}[A^{(\alpha)}_n]^\dagger_n=\hat{\pi}[A^{(\alpha)}_n]_n,\qquad
\hat{p}^\dagger_n=\hat{p}_n.
\end{equation}

The canonical phase space transformation induced by the gauge transformation in \textcolor{black}{$\varphi(q^n)$ of} the background gauge field is realised
on Hilbert space by the associated quantum operator
$U_\varphi(\hat{q}^n)=e^{\frac{i}{\hbar}\varphi(\hat{q}^n)}$, so that \textcolor{black}{the abstract canonical conjugate momenta operators
$\hat{\pi}[A^{(\alpha)}_n]_n$ transform covariantly},
\begin{equation}
\hat{\pi}[A^{(2)}_n]_n=\hat{\pi}[A^{(1)}_n]_n+\partial_n\varphi(\hat{q}^n)=U^\dagger_\varphi(\hat{q}^n)\,\hat{\pi}[A^{(1)}_n]_n\,U_\varphi(\hat{q}^n),\quad
\hat{\pi}[A^{(1)}_n]_n=U_\varphi(\hat{q}^n)\,\hat{\pi}[A^{(2)}_n]_n\,U^\dagger_\varphi(\hat{q}^n).
\end{equation}
However one must keep in mind that this operator $U_\varphi(\hat{q}^n)$ must act \textcolor{black}{on abstract quantum operators}
jointly with the gauge transformation of the vector gauge field,
$A^{(1)}_n(\hat{q}^n) \rightarrow A^{(1)}_n(\hat{q}^n)+\partial_n\varphi(\hat{q}^n)=A^{(2)}_n(\hat{q}^n)$, so that for instance
the \textcolor{black}{abstract} momenta operators $\hat{p}_n$ are \textcolor{black}{strictly gauge} invariant under such a combined transformation
\textcolor{black}{as it should}, namely,
\begin{equation}
\hat{p}_n=\hat{\pi}[A^{(2)}_n]_n-A^{(2)}_n(\hat{q}^n)=\hat{\pi}[A^{(1)}_n]_n-A^{(1)}_n(\hat{q}^n).
\end{equation}
Likewise the system's gauge invariant \textcolor{black}{abstract} quantum Hamiltonian (or for that matter, any gauge invariant \textcolor{black}{abstract}
quantum observable $\hat{\cal O}={\cal O}_0(\hat{q}^n,\hat{p}_n)$) is thus in the form of,
\begin{equation}
\hat{H}=\hat{H}_2=\hat{H}_1=H_\alpha(\hat{q}^n,\hat{\pi}[A^{(\alpha)}_n]_n)=H_0(\hat{q}^n,\hat{p}_n)
=H_0(\hat{q}^n,\hat{\pi}[A^{(\alpha)}_n]_n-A^{(\alpha)}_n(\hat{q}^n))=\hat{H}_0.
\end{equation}

In the notations of Section~\ref{Sect2}, the configuration space representations of the \textcolor{black}{abstract canonical}
conjugate momenta operators $\hat{\pi}[A^{(\alpha)}_n]_n$
given the configuration space eigenbases $|q^n;\alpha,\chi_\alpha\rangle$ with $\alpha=1,2$ are \textcolor{black}{expressed} as,
\begin{equation}
\langle q^n;\alpha,\chi_\alpha|\hat{\pi}[A^{(\alpha)}_n]_n|\psi\rangle = \frac{-i\hbar}{g^{1/4}(q^n)}\left(\partial_n+\frac{i}{\hbar}V^{(\alpha,\chi_\alpha)}_n(q^n)\right)
g^{1/4}(q^n)\, \langle q^n;\alpha,\chi_\alpha|\psi\rangle,\quad \alpha=1,2,
\end{equation}
where the vector fields $V^{(\alpha,\chi_\alpha)}_n(q^n)$ are related through the following complete U(1) gauge transformation inclusive of the changes
in the local quantum phase factors of \textcolor{black}{configuration space} eigenbases \textcolor{black}{of} quantum states,
\begin{equation}
V^{(2,\chi_2)}_n(q^n)=V^{(1,\chi_1)}_n(q^n)+\partial_n\left(\varphi(q^n)+\chi_{21}(q^n)+\chi_2(q^n)-\chi_1(q^n)\right).
\end{equation}
Likewise the \textcolor{black}{abstract} gauge invariant momenta operators $\hat{p}_n$ possess the following configuration space representations,
\begin{equation}
\langle q^n;\alpha,\chi_\alpha|\hat{p}_n|\psi\rangle = \frac{-i\hbar}{g^{1/4}(q^n)}
\left(\partial_n+\frac{i}{\hbar}\left(V^{(\alpha,\chi_\alpha)}_n(q^n)-A^{(\alpha)}_n(q^n)\right)\right) g^{1/4}(q^n)\,\langle q^n;\alpha,\chi_\alpha|\psi\rangle.
\end{equation}
In these last two matrix elements \textcolor{black}{for $\alpha=1,2$}, the configuration space wave functions of a same abstract quantum state $|\psi\rangle$ given the two choices of eigenbases
are related according to the phase transformation,
\begin{equation}
\langle q^n;2,\chi_2|\psi\rangle = e^{-\frac{i}{\hbar}(\chi_{21}(q^n) + \chi_2(q^n) - \chi_1(q^n))}\,\langle q^n;1,\chi_1|\psi\rangle.
\end{equation}
Note that the action of the abstract gauge invariant momenta operators $\hat{p}_n$ on any abstract quantum state $|\psi\rangle$ being given as,
\begin{equation}
\hat{p}_n|\psi\rangle = \int_M d^N\!q^n\sqrt{g(q^n)}\,|q^n;\alpha,\chi_\alpha\rangle \langle q^n;\alpha,\chi_\alpha|\hat{p}_n|\psi\rangle,\qquad \alpha=1,2,
\end{equation}
while the basis states $|q^n;\alpha,\chi_\alpha\rangle$ $(\alpha=1,2)$ are related by
\begin{equation}
|q^n;2,\chi_2\rangle = e^{\frac{i}{\hbar}(\chi_{21}(q^n)+\chi_2(q^n)-\chi_1(q^n))}\,|q^n;1,\chi_1\rangle,
\end{equation}
the \textcolor{black}{abstract} quantum operators $\hat{p}_n$ are indeed invariant under gauge transformations of the background vector gauge field $A^{(1)}_n(q^n)$.

To conclude, and specifically now in the case of a configuration space manifold of trivial first homotopy group such that all unitary
\textcolor{black}{configuration space} representations of the Heisenberg
algebra are unitarily equivalent, let us point to two noteworthy phase conventions for the choices of configuration space eigenstates,
such that the same choices of eigenbases $|q^n;1\rangle$ and $|q^n;2\rangle$ are effected \textcolor{black}{whether} before or after the gauge transformation
in $\partial_n\varphi(q^n)$ of the background vector gauge field $A^{(1)}_n(q^n)$, namely by choosing $\chi_1(q^n)=0=\chi_2(q^n)$.
Since the fundamental group of configuration space is trivial, any unitary \textcolor{black}{configuration space}
representation of the Heisenberg algebra is labelled by a flat U(1) gauge connection which
is pure gauge, $V^{(1)}_n(q^n)=\partial_nV^{(1)}(q^n)$. By an appropriate choice of local quantum phase for the basis states $|q^n;1\rangle$,
it is then always possible to also choose the canonical representation of the Heisenberg algebra such that $V^{(1)}_n(q^n)=0$.

Given this preliminary set of phase conventions, the first noteworthy choice for the remaining freedom in the choice of $\chi_{21}(q^n)$ is such that the
two eigenbases are identical, $|q^n;2\rangle=|q^n;1\rangle \equiv |q^n\rangle$, namely such that $\chi_{21}(q^n)=0$, which therefore implies
\begin{equation}
\psi_{(2,\chi_2)}(q^n)=\psi_{(2)}(q^n)=\psi_{(1)}(q^n)=\psi_{(1,\chi_1)}(q^n) \equiv \psi(q^n).
\end{equation}
All the phase choices having been specified with $\chi_1(q^n)=\chi_2(q^n)=\chi_{21}(q^n)=0$ and $V^{(1)}_n(q^n)=0$,
the gauge transformation in the background vector gauge field induces a transformation of the initial canonical unitary \textcolor{black}{configuration space}
representation of the Heisenberg algebra into the \textcolor{black}{now}
non canonical unitary \textcolor{black}{configuration space} representation which is labelled by the flat but nonvanishing pure gauge U(1) connection
\begin{equation}
V^{(2)}_n(q^n)=\partial_n\varphi(q^n).
\end{equation}
Consequently under such specific conditions the configuration space representation of the gauge invariant quantum momenta $\hat{p}_n$ is then provided by,
\begin{equation}
\langle q^n|\hat{p}_n|\psi\rangle  = \frac{-i\hbar}{g^{1/4}(q^n)}\left(\partial_n+\frac{i}{\hbar}\left(-A^{(1)}_n(q^n)\right)\right)\,
g^{1/4}(q^n)\,\langle q^n|\psi\rangle.
\end{equation}

On the other hand, still given the same preliminary set of phase conventions such that $\chi_1(q^n)=0=\chi_2(q^n)$ and $V^{(1)}_n(q^n)=0$,
since the transformed flat U(1) connection $V^{(2)}_n(q^n)$ is pure gauge---all unitary Heisenberg algebra representations being unitarily equivalent to the
canonical one when the fundamental group of the configuration space is trivial---one may always assume a choice for the local quantum phase
of the basis states $|q^n;2\rangle$ such that $V^{(2)}_n(q^n)=0$. Thus the second noteworthy choice for the remaining freedom in the choice, this time,
of the unitary representation of the Heisenberg algebra to be used once the gauge transformation of the background vector gauge field is effected,
is such that $V^{(2)}_n(q^n)=0$. Consequently by requiring that the unitary Heisenberg algebra representations remains canonical
with $V^{(\alpha)}_n(q^n)=0$ whether $\alpha=1,2$, one must choose the $|q^n;2\rangle$ basis such that $\chi_{21}(q^n)=-\varphi(q^n)$, namely,
\begin{equation}
|q^n;2\rangle = e^{-\frac{i}{\hbar}\varphi(q^n)}\,|q^n;1\rangle,\qquad
\langle q^n;2|\psi\rangle = e^{\frac{i}{\hbar}\varphi(q^n)}\,\langle q^n;1|\psi\rangle,\qquad
\psi_{(2)}(q^n)=e^{\frac{i}{\hbar}\varphi(q^n)}\,\psi_{(1)}(q^n),
\label{eq:gauge2}
\end{equation}
while the abstract quantum momenta $\hat{p}_n$ then have the following two different configuration space wave function representations,
\begin{equation}
\langle q^n;1|\hat{p}_n|\psi\rangle = \frac{-i\hbar}{g^{1/4}(q^n)}\left(\partial_n+\frac{i}{\hbar}\left(-A^{(1)}_n(q^n)\right)\right)\,g^{1/4}(q^n)\,\langle q^n;1|\psi\rangle,
\end{equation}
\begin{equation}
\langle q^n;2|\hat{p}_n|\psi\rangle = \frac{-i\hbar}{g^{1/4}(q^n)}\left(\partial_n+\frac{i}{\hbar}\left(-A^{(1)}_n(q^n)-\partial_n\varphi(q^n)\right)\right)\,
g^{1/4}(q^n)\,\langle q^n;2|\psi\rangle,
\end{equation}
which are gauge covariant nevertheless for gauge transformations of the background vector gauge field on account of (\ref{eq:gauge2})\textcolor{black}{,
namely the fact that the local phases of the configuration space eigenstates $|q^n;2\rangle$ and $|q^n;1\rangle$ differ by the factor
$e^{-\frac{i}{\hbar}\varphi(q^n)}$ associated to these gauge transformations given the present choice of all necessary phase conventions}
(while the abstract operators $\hat{p}_n$ themselves are indeed exactly gauge invariant\textcolor{black}{, as are the abstract quantum states $|\psi\rangle$ as well}).

To paraphrase the end of Section~\ref{Sect2} as a conclusion to the present one, applying a gauge transformation to a background vector gauge field
to which a dynamics is coupled (whether for a mechanical system or a field theory) thus certainly has subtle and non trivial
consequences for the quantised system and specifically for the unitary representations to be chosen for its Heisenberg algebra,
that must properly be accounted for to ascertain a strictly identical physical and gauge invariant quantum dynamics.

\section{The Landau Problem---Canonical Formulation}
\label{Sect4}

Finally, let us apply the general considerations of Sections~\ref{Sect2} and \ref{Sect3} \textcolor{black}{specifically} to the Landau problem,
in the notations \textcolor{black}{of} the Introduction. The local action principle is then given by,
\begin{equation}
S[x_i;A_i]=\int dt\, L(x_i,\dot{x}_i;A_i),\qquad
L(x_i,\dot{x}_i;A_i) = \frac{1}{2}m \dot{x}^2_i + q \dot{x}_i A_i(x_i),\qquad i=1,2,
\end{equation}
where $\vec{A}(\vec{x}\,)=A_i(x_i)\,\hat{e}_i$ is a particular gauge choice for the vector potential from which the static and uniform
magnetic field derives, $\vec{B}=\vec{\nabla}\times\vec{A}(\vec{x}\,)=B\,\hat{e}_3$.

To allow for a discussion as general as possible, the parametrisation to be used for the background vector gauge potential is in the form of
\begin{equation}
A_i(x_i) = -\frac{1}{2} B \epsilon_{ij} (x_j-x_{0j})+\frac{\partial}{\partial x_i}\bar{\varphi}(x_i),\quad
\bar{\varphi}(x_i)=-\frac{1}{2}\alpha \, B\,(x_1-x_{01})(x_2- x_{02})\,+\,\varphi(x_i),
\label{eq:A1}
\end{equation}
or in terms of its cartesian components,
\begin{equation}
A_1(x_i)=-\frac{1}{2}(\alpha+1)B(x_2-x_{02}) + \partial_1\varphi(x_i),\quad
A_2(x_i) = -\frac{1}{2}(\alpha-1) B (x_1 -x _{01}) + \partial_2\varphi(x_i).
\label{eq:A2}
\end{equation}
In (\ref{eq:A1}) the function $\bar{\varphi}(\vec{x}\,)$ represents any arbitrary gauge transformation of the vector potential away from the symmetric gauge
centered at $\vec{x}=\vec{x}_0$, \textcolor{black}{the latter being} defined by the contribution in $-B\epsilon_{ij}(x_j-x_{0j})/2$,
namely $\vec{B}\times(\vec{x}-\vec{x}_0)/2$,
to $A_i(x_i)$. In turn the function $\bar{\varphi}(\vec{x}\,)$
is represented in terms of another arbitrary such function $\varphi(\vec{x}\,)$, combined with the contribution in $-\alpha B(x_1-x_{01})(x_2-x_{02})/2$
involving the arbitrary real parameter $\alpha\in\mathbb{R}$

This form of general parametrisation for the choice of the background vector gauge potential $\vec{A}(\vec{x}\,)$, dubbed the $(\alpha,\varphi(\vec{x}\,))$ gauge
hereafter, is motivated by the following observations. For the choice $(\alpha=0,\varphi(\vec{x}\,)=0)$ one has the usual symmetric gauge (centered at $\vec{x}_0$).
When $(\alpha=1,\varphi(\vec{x}\,)=0)$, one has the first Landau gauge which is invariant under translations in $x_1$. And for the choice
$(\alpha=-1,\varphi(\vec{x}\,)=0)$, one has the second Landau gauge which is invariant under translations in $x_2$. These are the three standard choices
of gauges for the Landau problem which have been considered time and again in the literature. Herein the most general gauge choice possible
for the background vector gauge potential is \textcolor{black}{thus} considered, with these three gauge choices as particular cases determined by specific values for
the free parameter $\alpha$ and the free function $\varphi(\vec{x}\,)$.

The Lagrange function of the system is therefore given as,
\begin{eqnarray}
&& L(x_i,\dot{x}_i;A_i) = \frac{1}{2}m\dot{x}^2_i-\frac{1}{2}qB\epsilon_{ij}\dot{x}_i(x_j-x_{0j})+\frac{d}{dt}\bar{\varphi}(x_i) \nonumber \\
&& = \frac{1}{2}m\dot{x}^2_i-\frac{1}{2}(\alpha+1)qB\dot{x}_1(x_2-x_{02})-\frac{1}{2}(\alpha-1)qB\dot{x}_2(x_1-x_{01})+\frac{d}{dt}\varphi(x_i).
\label{eq:L-Landau}
\end{eqnarray}
Incidentally note how the corresponding action, $S[x_i;A_i]=\int dt L(x_i,\dot{x}_i;A_i)$, is manifestly invariant under constant translations in time,
while it is invariant up to some total divergence in time under constant spatial translations in the plane, \textcolor{black}{as well as under}
constant planar rotations centered at the
point of position vector $\vec{x}_0$. Consequently Noether's first theorem \textcolor{black}{(see for example Ref.\cite{Gov1})}
guarantees the existence of four conserved quantities or Noether charges
as the generators of these global symmetries of the Landau problem, independently of the choice of gauge. Namely the total energy, the two generators
for spatial translations, and the generator for rotations about $\vec{x}_0$, denoted already in the Introduction as $E$, $T_i$ ($i=1,2)$, and $M_3$, respectively.
Furthermore as physical observables, these quantities need to be invariant under gauge transformations of the background vector gauge potential
$\vec{A}(\vec{x}\,)$.

In its canonical Hamiltonian formulation, the system's symplectic phase space is spanned by the coordinates $x_i$ as well as their \textcolor{black}{canonical}
conjugate momenta $\pi[\vec{A}\,]_i$ given as
\begin{equation}
\pi[\vec{A}\,]_i=\frac{\partial L}{\partial\dot{x}_i}=m\dot{x}_i+q A_i(x_i)= p_i + A_i(x_i),\qquad
p_i=\pi[\vec{A}\,]_i-qA_i(x_i),
\end{equation}
where the momenta $p_i=m\dot{x_i}$ are the cartesian components of the velocity momentum of the particle of mass $m$. For the reasons
discussed in Section~\ref{Sect3}, $\pi[\vec{A}\,]_i$ are dependent on the choice of vector gauge potential $\vec{A}(\vec{x}\,)$ and are not invariant
under gauge transformations of the latter field. In contradistinction the velocity momentum $\vec{p}=p_i\,\hat{e}_i$ is independent of $\vec{A}(\vec{x}\,)$
and is \textcolor{black}{strictly} invariant under gauge transformations of that background gauge field. Furthermore such gauge transformations of $\vec{A}(\vec{x}\,)$
induce canonical transformations of the conjugate momenta $\pi[\vec{A}\,]_i$ in phase space \cite{Gov3}.

The phase space coordinates $(x_i,\pi[\vec{A}\,]_i)$ are canonical with canonical values for their Poisson brackets,
\begin{equation}
\left\{x_i,x_j\right\}=0,\qquad
\left\{x_i,\pi[\vec{A}\,]_j\right\}=\delta_{ij},\qquad
\left\{\pi[\vec{A}\,]_i,\pi[\vec{A}\,]_j\right\}=0,
\end{equation}
while the Poisson brackets of the gauge invariant phase space parametrisation $(x_i,p_i)$ take the values,
\begin{equation}
\left\{x_i,x_j\right\}=0,\qquad
\left\{x_i,p_j\right\}=\delta_{ij},\qquad
\left\{p_i,p_j\right\}=qB\,\epsilon_{ij}.
\end{equation}
The gauge invariant Hamiltonian also reads,
\begin{equation}
H(x_i,\pi[\vec{A}\,]_i)=\dot{x}_i \pi[\vec{A}\,]_i - L(x_i,\dot{x}_i)=\frac{1}{2m}\left(\vec{\pi}[\vec{A}\,]-q\vec{A}(x_i)\right)^2=\frac{1}{2m} p^2_i=H_0(p_i),
\end{equation}
and generates the following first-order Hamiltonian equations of motion,
\begin{equation}
\dot{x}_i=\frac{1}{m}\left(\pi[\vec{A}\,]_i-qA_i(x_i)\right)=\frac{1}{m}p_i,\quad
\dot{\pi}[\vec{A}\,]_i=\frac{q}{m}\partial_iA_j(x_i)\left(\pi[\vec{A}\,]_j-qA_j(x_i)\right)=\frac{q}{m}\partial_iA_j(x_i) p_j,
\end{equation}
so that,
\begin{equation}
m\ddot{x}_i=q\partial_iA_j(x_i)\dot{x}_j -q\partial_j A_i(x_i) \dot{x}_j=qB\epsilon_{ij}\dot{x}_j,\qquad
\dot{p}_i=\frac{qB}{m}\epsilon_{ij} p_j.
\end{equation}

From a complementary point of view, note that an action of the form $\int dt(m\dot{x}^2_i/2)$ is equivalent to the following one in which
auxiliary gaussian variables $p_i$ are introduced,
\begin{equation}
\int dt(\dot{x}_i p_i - p^2_i/(2m))=\int dt\left(-\frac{1}{2m}(p_i-m\dot{x}_i)^2+\frac{1}{2}m\dot{x}^2_i\right),
\end{equation}
whose variational equations then imply $p_i=m\dot{x}_i$.
Namely $p_i$ are indeed the components of the velocity momentum. Consequently the action in (\ref{eq:L-Landau}) is equivalent to the following
first-order one, $S_H[x_i,p_i;A_i]=\int dt L_H$ with
\begin{equation}
L_H=\dot{x}_i p_i - \frac{1}{2m}p^2_i+q \dot{x}_i A_i(x_i)=\dot{x}_i(p_i+qA_i(x_i))-\frac{1}{2m}p^2_i=\dot{x}_i \omega_i(x_i,p_i) - H_0(p_i),
\end{equation}
as well as
\begin{equation}
\omega_i(x_i,p_i)=p_i+qA_i(x_i).
\end{equation}
In otherwords, $S_H[x_i,p_i;A_i]$ is the Hamiltonian first-order action of the system, with the phase space symplectic 1-form
\begin{equation}
\omega=dx_i\, \omega_i=dx_i \,(p_i+qA_i(x_i)).
\end{equation}
Indeed, it may be checked that the inverse of the symplectic 2-form $\Omega=d\wedge\omega=\frac{1}{2}dx_i\wedge dx_j \Omega_{ij}$
coincides with the above Poisson brackets for the phase space coordinates $(x_i,p_i)$, as it should.

By considering the variation of the Lagrange function $L(x_i,\dot{x}_i;A_i)$ under infinitesimal transformations of the global symmetries listed above,
the first Noether identity \cite{Gov1} leads to the identification of the associated conserved Noether charges as follows,
\begin{eqnarray}
E &=& \frac{1}{2m}p^2_i=\frac{1}{2m}\left(\pi[\vec{A}\,]_i-qA_i(x_i)\right)^2, \nonumber \\
T_i &=& p_i-qB\epsilon_{ij}(x_j-x_{0j})=\pi[\vec{A}\,]_i - qA_i(x_i) -qB\epsilon_{ij}(x_j - x_{0j}),  \\
M_3 &=& \epsilon_{ij}(x_i- x_{0i})p_j + \frac{1}{2}qB(x_i-x_{0i})^2=\epsilon_{ij}(x_i-x_{0i})(\pi[\vec{A}\,]_j-qA_j(x_i))+\frac{1}{2}qB(x_i-x_{0i})^2, \nonumber 
\end{eqnarray}
which in particular obey the relation
\begin{equation}
\vec{T}\,^2=2mE + 2qB M_3=2mE+2s m\omega_c M_3.
\end{equation}
Note well that all these quantities are expressed solely in terms of $(x_i,p_i)$, and are thus indeed \textcolor{black}{strictly}
invariant under gauge transformations of the vector gauge
potential, thereby determining physical observables of the system. In particular the quantities $\epsilon_{ij} T_j/qB+ x_{0i}$ define the cartesian components
of the magnetic centre position vector, $x_{ci}$\textcolor{black}{, which are thus indeed strictly gauge invariant observables as well}.

These \textcolor{black}{strictly gauge invariant} Noether charges possess the following closed algebra of Poisson brackets,
\begin{equation}
\left\{T_i,H\right\}=0,\quad \left\{M_3,H\right\}=0,\quad
\left\{T_i,T_j\right\}=-qB\epsilon_{ij},\quad
\left\{T_i,M_3\right\}=-\epsilon_{ij} T_j,
\end{equation}
which implies for the magnetic centre components,
\begin{equation}
\left\{x_{ci},x_{cj}\right\}=-\frac{1}{qB}\,\epsilon_{ij},\qquad x_{ci}=x_{0i}+\frac{1}{qB}\epsilon_{ij} T_j.
\end{equation}
Infinitesimal spatial translations are indeed generated by the Noether charges $T_i$, since
\begin{equation}
\delta_{\vec{a}} x_i = \left\{x_i,a_j T_j\right\} =a_i,\qquad
\delta_{\vec{a}} p_i = \left\{p_i,a_j T_j\right\} = 0,
\end{equation}
and likewise for spatial rotations about $\vec{x}_0$ and the Noether charge $M_3$,
\begin{equation}
\delta_\theta x_i = \left\{x_i,\theta M_3\right\} =-\theta\,\epsilon_{ij}\,(x_j-x_{0j}),\qquad
\delta_\theta p_i = \left\{p_i,\theta M_3\right\} =-\theta\,\epsilon_{ij}\,p_j.
\end{equation}

By evaluating the explicit expressions for \textcolor{black}{the gauge invariant observables} $T_i$ and $M_3$
\textcolor{black}{in terms of $(x_i,\pi[\vec{A}\,]_i)$} in the $(\alpha,\varphi(\vec{x}\,))$ gauge, one finds
\begin{equation}
T_1=p_1-qB(x_2-x_{02})=\pi[\vec{A}\,]_1-\left(-\frac{1}{2}(\alpha-1)qB(x_2-x_{02})+\partial_1(q\varphi(x_i))\right),
\end{equation}
\begin{equation}
T_2=p_2+qB(x_1-x_{01})=\pi[\vec{A}\,]_2-\left(-\frac{1}{2}(\alpha+1)qB(x_1-x_{01})+\partial_2(q\varphi(x_i))\right),
\end{equation}
and,
\begin{eqnarray}
M_3 &=& \ \ \ \left[(x_1-x_{01})\pi[\vec{A}\,]_2 - (x_2-x_{02})\pi[\vec{A}\,]_1\right]  \\
&& +\left[\frac{1}{2}\alpha\, qB\left((x_1-x_{01})^2-(x_2-x_{02})^2\right)-(x_1-x_{01})\partial_2(q\varphi(x_i))+(x_2-x_{02})\partial_1(q\varphi(x_i))\right]. \nonumber
\end{eqnarray}
Consequently the first Landau gauge $(\alpha=1,\varphi(\vec{x}\,)=0)$ is distinguished by the fact that the \textcolor{black}{gauge invariant}
Noether charge $T_1$ and generator of translations in $x_1$ then coincides precisely with the \textcolor{black}{gauge variant}
conjugate momentum $\pi[\vec{A}\,]_1$. Likewise for the second Landau gauge $(\alpha=-1,\varphi(\vec{x}\,)=0)$
which is such that the \textcolor{black}{gauge invariant} Noether charge $T_2$ and generator of translations in $x_2$ then coincides precisely with the
\textcolor{black}{gauge variant} conjugate momentum $\pi[\vec{A}\,]_2$.
And finally the symmetric gauge $(\alpha=0,\varphi(x_i)=0)$ is distinguished by the fact that the \textcolor{black}{gauge invariant}
Noether charge $M_3$ and generator of rotations around
$\vec{x}_0$ then coincides precisely with the \textcolor{black}{gauge variant} canonical angular-momentum defined by the \textcolor{black}{canonical} conjugate momenta,
$(\vec{x}-\vec{x}_0)\times\vec{\pi}[\vec{A}\,]=L^c_3[\vec{A}\,]\,\hat{e}_3$. These observations explain the simplifications\textcolor{black}{---implicitly largely
exploited in the vast Landau problem literature---}implied by these specific gauge choices
when solving the dynamics of the Landau problem, certainly within the quantum context when wanting to diagonalise the Hamiltonian jointly with any one of the
three \textcolor{black}{gauge invariant} conserved Noether charges $T_i$ ($i=1,2)$ or $M_3$.

In the absence of the applied magnetic field $\vec{B}$, all three quantities, namely the \textcolor{black}{canonical}
conjugate momentum $\vec{\pi}[\vec{A}\,]$, the velocity momentum
$\vec{p}$, and the generator for spatial translations $\vec{T}$, reduce to the same physical observable, namely the velocity momentum, $\vec{p}$. In the presence
of the applied magnetic field however, these three vector quantities are distinct, with only two of these being gauge invariant and physical observables
for the Landau problem, \textcolor{black}{namely $\vec{T}$ and $\vec{p}$}, with the relations,
\begin{equation}
\vec{T}=\vec{p}-q(\vec{x}-\vec{x}_0)\times\vec{B}=\vec{\pi}[\vec{A}\,]-q(\vec{x}-\vec{x}_0)\times\vec{B}-q\vec{A}(\vec{x}\,),\qquad
\vec{p}=\vec{\pi}[\vec{A}\,]-q\vec{A}(\vec{x}\,).
\label{eq:Relation1}
\end{equation}
Likewise for what concerns properties of the Landau problem under rotations in the plane, besides the \textcolor{black}{gauge invariant}
generator of such transformations, \textcolor{black}{namely} $M_3$,
one may also consider the canonical angular-momentum $L^c_3[\vec{A}\,]$, and the velocity or orbital angular-momentum $L_3$, defined as,
\begin{equation}
L^c_3[\vec{A}\,]=\epsilon_{ij }(x_i - x_{0i})\,\pi[\vec{A}\,]_j,\qquad
L_3=\epsilon_{ij} (x_i - x_{0i})\,p_j,
\end{equation}
which again in the absence of the magnetic field all reduce to the orbital angular-momentum $(\vec{x}-\vec{x}_0)\times\vec{p}=L_3\,\hat{e}_3$.
In the presence of the magnetic field however, only $L_3$ and $M_3$ are genuine gauge invariant and physical observables\textcolor{black}{, while
$L^c_3[\vec{A}\,]$ is gauge variant and thus not physical}. Yet all three quantities
share the following relations,
\begin{equation}
L_3=\epsilon_{ij}(x_i-x_{0i})\,T_j - qB(\vec{x}-\vec{x}_0)^2=L^c_3[\vec{A}\,]-\frac{1}{2}qB(\vec{x}-\vec{x}_0)^2-\epsilon_{ij}(x_i-x_{0i})\,\partial_j(q\bar{\varphi}(x_i)),
\label{eq:Relation2}
\end{equation}
as well as,
\begin{equation}
M_3=L_3+\frac{1}{2}qB(\vec{x}-\vec{x}_0)^2=L^c_3[\vec{A}\,]-\epsilon_{ij}(x_i-x_{0i})\,\partial_j(q\bar{\varphi}(x_i)).
\label{eq:Relation3}
\end{equation}
Since the relations (\ref{eq:Relation1},\ref{eq:Relation2},\ref{eq:Relation3}) are linear and quadratic in phase space coordinates that commute with one another
for the quantised system, these relations remain valid at the quantum level as well, thereby translating into likewise relations for the matrix elements
of these quantities for whatever abstract quantum states.

\section{The Landau Problem---Canonical Quantisation}
\label{Sect5}

Given the general $(\alpha,\varphi(\vec{x}\,))$ gauge \textcolor{black}{fixing} choice, the quantum Landau problem is defined by the \textcolor{black}{following}
Heisenberg algebra commutation relations (considered at some implicit reference time $t=t_0$),
\begin{equation}
\left[\hat{x}_i,\hat{x}_j\right]=0,\qquad
\left[\hat{x}_i,\pi[\vec{A}\,]_j\right]=i\hbar\,\delta_{ij}\,\mathbb{I},\qquad
\left[\hat{\pi}[\vec{A}\,]_i,\hat{\pi}[\vec{A}\,]_j\right]=0,
\end{equation}
or \textcolor{black}{equivalently},
\begin{equation}
\left[\hat{x}_i,\hat{x}_j\right]=0,\qquad
\left[\hat{x}_i,\hat{p}_j\right]=i\hbar\,\delta_{ij}\,\mathbb{I},\qquad
\left[\hat{p}_i,\hat{p}_j\right]=i\hbar\,qB\,\epsilon_{ij}\,\mathbb{I},
\end{equation}
with the hermiticity requirements $\hat{x}^\dagger_i=\hat{x}_i$, $\hat{\pi}[\vec{A}\,]^\dagger_i=\hat{\pi}[\vec{A}\,]_i$ and $\hat{p}^\dagger_i=\hat{p}_i$,
as well as the \textcolor{black}{abstract gauge invariant} quantum Hamiltonian,
\begin{equation}
\hat{H}=\frac{1}{2m}\left(\hat{\pi}[\vec{A}\,]_i - q A_i(\hat{x}_i)\right)^2=\frac{1}{2m}\hat{p}^2_i.
\end{equation}
As quantum operators the \textcolor{black}{gauge invariant} Noether charges now read,
\begin{equation}
\hat{E} =\hat{H} = \frac{1}{2m}\hat{p}^2_i,\quad
\hat{T}_i = \hat{p}_i-qB\epsilon_{ij}(\hat{x}_j-x_{0j}\mathbb{I}),\quad
\hat{M}_3 = \epsilon_{ij}(\hat{x}_i- x_{0i}\mathbb{I})\hat{p}_j + \frac{1}{2}qB(\hat{x}_i-x_{0i}\mathbb{I})^2,
\end{equation}
with the algebra of commutation relations,
\begin{equation}
\left[\hat{T}_i,\hat{H}\right]=0=\left[\hat{M}_3,\hat{H}\right],\qquad
\left[\hat{T}_1,\hat{T}_2\right]=-i\hbar\,qB\,\mathbb{I}=-i\hbar\, sm\omega_c\,\mathbb{I},\qquad
\left[\hat{T}_i,\hat{M}_3\right]=-i\hbar\,\epsilon_{ij}\,\hat{T}_j.
\end{equation}
In particular note well that the \textcolor{black}{gauge invariant} magnetic centre coordinates are such that
\begin{equation}
\left[\hat{x}_{ci},\hat{x}_{cj}\right]=-\frac{i\hbar}{qB}\epsilon_{ij}\,\mathbb{I},\qquad
\hat{x}_{ci}=x_{0i}\,\mathbb{I} + \frac{1}{qB}\epsilon_{ij}\,\hat{T}_j.
\end{equation}
Therefore as gauge invariant and physical observables, the coordinate operators $\hat{x}_{ci}$ define a noncommutative geometry in the plane
(as do the translation generators $\hat{T}_i$).

The following commutation relations also apply,
given that $\hat{L}_3=\hat{M}_3-\frac{1}{2}qB(\hat{\vec{x}}-\vec{x}_0\mathbb{I})^2$,
\begin{equation}
\left[\hat{x}_{ci},\hat{H}\right]=0,\qquad
\left[\hat{L}_3,\hat{M}_3\right]=0,\qquad
\left[\hat{L}_3,\hat{H}\right]=-\frac{1}{2}i\hbar\,s\omega_c\,\left[(\hat{x}_i-x_{0i}\mathbb{I})\hat{p}_i+\hat{p}_i(\hat{x}_i-x_{0i}\mathbb{I})\right],
\end{equation}
showing  that the magnetic centre position vector operator $\hat{\vec{x}}_c$ is a conserved gauge invariant physical quantity, while $\hat{L}_3$,
even though not conserved yet gauge invariant and \textcolor{black}{thus} physical, is invariant under rotations about $\vec{x}_0$.
Finally one also finds, for infinitesimal spatial symmetry transformations,
\begin{equation}
\left[\hat{x}_i,\hat{T}_j\right]=i\hbar\,\delta_{ij}\,\mathbb{I},\quad
\left[\hat{p}_i,\hat{T}_j\right]=0,\qquad
\left[\hat{x}_i-x_{0i}\mathbb{I},\hat{M}_3\right]=-i\hbar\,\epsilon_{ij}(\hat{x}_j-x_{0j}\mathbb{I}),\quad
\left[\hat{p}_i,\hat{M}_3\right]=-i\hbar\,\epsilon_{ij}\hat{p}_j,
\end{equation}
while,
\begin{equation}
\left[\hat{x}_{ci},\hat{x}_{cj}\right]=-\frac{i\hbar}{qB}\epsilon_{ij}\,\mathbb{I},\qquad
\left[\hat{p}_i,\hat{p}_j\right]=i\hbar\,qB\,\epsilon_{ij}\,\mathbb{I},\qquad
\left[\hat{x}_{ci},\hat{p}_j\right]=0.
\label{eq:triple}
\end{equation}
The last three sets of commutation relations thus show that each of the pairs of quantum operators, $(\hat{x}_{c1},\hat{x}_{c2})$ and $(\hat{p}_1,\hat{p}_2)$,
defines on its own again a Heisenberg algebra, while these two Heisenberg algebras commute with one another, thus sharing these same properties with
the pairs $(\hat{x}_1,\hat{\pi}[\vec{A}\,]_1)$ and $(\hat{x}_2,\hat{\pi}[\vec{A}\,]_2)$. In contradistinction to the latter two pairs however, the pairs $\hat{x}_{ci}$
and $\hat{p}_i$ are each gauge invariant and physical, with the simple decoupled \textcolor{black}{quantum Hamiltonian} equations of motion,
\begin{equation}
i\hbar\frac{d}{dt}\hat{x}_{ci}=\left[\hat{x}_{ci},\hat{H}\right]=0,\qquad
i\hbar\frac{d}{dt}\hat{p}_i=\left[\hat{p}_i,\hat{H}\right]=i\, s\hbar\omega_c\,\epsilon_{ij}\, \hat{p}_j.
\end{equation}
Note well that the \textcolor{black}{gauge invariant}
magnetic centre component operators, $\hat{x}_{ci}$, commute with the quantum Hamiltonian. Hence, these operators (or equivalently
the Noether generators for spatial translations, $\hat{T}_i$) act by mapping \textcolor{black}{between}
quantum states within a same Landau level of definite energy eigenvalue,
\textcolor{black}{that} differ in their magnetic centre (or total angular-momentum) probability distribution. On the other hand,
the complementary set of \textcolor{black}{gauge invariant}
quantum operators, namely the components of the velocity momentum, $\hat{p}_i$, necessarily map between quantum states
that belong to different Landau levels whose energies differ by a single quantum of energy whose gap is set by the cyclotron frequency, namely $\hbar\omega_c$.

In wanting to identify the physical \textcolor{black}{and gauge invariant}
spectral content of the quantum Landau problem, one needs to consider a complete basis of eigenstates of a maximal
set of commuting quantum operators which includes the \textcolor{black}{gauge invariant} quantum Hamiltonian $\hat{H}$.
The system being that of two degrees of freedom, $\hat{x}_i$,
a second operator that commutes with $\hat{H}$ is required, namely a conserved gauge invariant and physical observable. Thus one may choose
to work in a basis of states which diagonalises either the pair $(\hat{T}_1,\hat{H})$, or the pair $(\hat{T}_2,\hat{H})$, or the pair $(\hat{M}_3,\hat{H})$,
and this still given the general choice of $(\alpha,\varphi(\vec{x}\,))$ gauge without the need to restrict to any one of the distinguished Landau or symmetric
gauges.

When diagonalising any such pair of physical observables, one also needs to specify a choice of unitary \textcolor{black}{configuration space}
representation of the Heisenberg algebra.
Irrespective of the values for the gauge fixing parameters of the $(\alpha,\varphi(\vec{x}\,))$ gauge, for simplicity
let us make the choice of the canonical configuration space wave function representation $\psi(\vec{x}\,)=\langle\vec{x}\,|\psi\rangle$
with thus the following correspondence rules when acting on an abstract quantum state\footnote{Since the two dimensional plane defines a flat Riemannian manifold
with trivial Euclidean metric $g_{ij}(x_i)=\delta_{ij}$, the volume factor $\sqrt{g(x_i)}$ of the general discussion of Section~\ref{Sect2} takes the value $g(x_i)=1$.}
$|\psi\rangle$,
\begin{equation}
\hat{x}_i \rightarrow x_i;\qquad
\hat{\pi}[\vec{A}\,]_i \rightarrow  -i\hbar\frac{\partial}{\partial x_i};\qquad
\hat{p}_i \rightarrow -i\hbar\frac{\partial}{\partial x_i} - q A_i(x_i).
\end{equation}
However let us emphasize that when considering the same canonical unitary representation of the Heisenberg algebra for another choice of values for
gauge fixing parameters, say $(\alpha_2,\varphi_2(\vec{x}\,))$, which amounts to a gauge transformation---from the gauge with values
$(\alpha,\varphi(\vec{x}\,))$---defined by the function $\Delta\bar{\varphi}(\vec{x}\,)$ given as
\begin{equation}
\Delta\bar{\varphi}(\vec{x}\,)=\bar{\varphi}_2(\vec{x}\,)-\bar{\varphi}(\vec{x}\,)
=-\frac{1}{2}(\alpha_2-\alpha) B (x_1-x_{01})(x_2-x_{02})+\varphi_2(\vec{x}\,)-\varphi(\vec{x}\,),
\label{eq:Delta}
\end{equation}
and such that $\vec{A}(\vec{x}\,) \equiv \vec{A}_{(\alpha,\varphi)}(\vec{x}\,)
\rightarrow \vec{A}_{(\alpha_2,\varphi_2)}(\vec{x}\,)=\vec{A}(\vec{x})+\vec{\nabla}\Delta\bar{\varphi}(\vec{x}\,)$,
then according to the general discussion presented at the very end of Section~\ref{Sect3} for such a choice of canonical unitary 
representation of the Heisenberg algebra, a change of the local quantum phase of the position eigenstates must be effected concurrently as well, such that,
\begin{equation}
|\vec{x};2\rangle=e^{-\frac{i}{\hbar}q\Delta\bar{\varphi}(\vec{x}\,)}\,|\vec{x}\,\rangle,\qquad
\psi_{(2)}(\vec{x}\,)=\langle \vec{x};2|\psi\rangle = e^{\frac{i}{\hbar}q\Delta\bar{\varphi}(\vec{x}\,)}\, \langle\vec{x}|\psi\rangle=
e^{\frac{i}{\hbar}q\Delta\bar{\varphi}(\vec{x}\,)}\,\psi(\vec{x}\,).
\label{eq:psi}
\end{equation}
\textcolor{black}{Note well that abstract quantum states $|\psi\rangle$ are strictly invariant under the gauge transformation $\Delta\bar{\varphi}(\vec{x}\,)$,
while their wave function representations then indeed transform contravariantly with the above phase factor simply because of the induced
{\sl passive} unitary U(1) transformation of the configuration space eigenbasis $|\vec{x}\,\rangle$ being mapped to the new gauge rotated such eigenbasis
$|\vec{x}; 2\rangle$, as is implied by the choice of canonical unitary configuration space representation of the Heisenberg algebra ({\sl i.e.}, that
with a vanishing flat U(1) connection) being made irrespective of the choice of gauge for the background vector field.}

Given the \textcolor{black}{thereby specified choices and phase conventions for the} canonical configuration space wave function representation
of the Heisenberg algebra for any specific choice of gauge fixing parameters $(\alpha,\varphi(\vec{x}\,))$, the \textcolor{black}{abstract gauge invariant}
quantum Hamiltonian operator is realised by the differential operator,
\begin{equation}
\hat{H}:\ \ \ -\frac{\hbar^2}{2m}\left(\partial_1-\frac{i}{\hbar}q A_1(x_i)\right)^2 - \frac{\hbar^2}{2m}\left(\partial_2-\frac{i}{\hbar} qA_2(x_i)\right)^2,
\end{equation}
in which the components of the vector gauge potential are given as,
\begin{eqnarray}
qA_1(x_i) &=& -\frac{1}{2}qB(x_2-x_{02})+\partial_1(q\bar{\varphi}(x_i))= -\frac{1}{2}(\alpha+1)qB(x_2-x_{02})+\partial_1(q\varphi(x_i)), \nonumber \\
qA_2(x_i) &=& \frac{1}{2}qB(x_1-x_{01})+\partial_2(q\bar{\varphi}(x_i)) = -\frac{1}{2}(\alpha-1)qB(x_1-x_{01})+\partial_2(q\varphi(x_i)).
\end{eqnarray}
Likewise for the other \textcolor{black}{gauge invariant} Noether charges, one finds \textcolor{black}{for $\hat{T}_i$ ($i=1,2$)},
\begin{equation}
\hat{T}_1:\ \ -i\hbar\partial_1-\left(-\frac{1}{2}(\alpha-1)qB(x_2-x_{02})+\partial_1 (q\varphi(x_i))\right),
\end{equation}
\begin{equation}
\hat{T}_2:\ \ -i\hbar\partial_2-\left(-\frac{1}{2}(\alpha+1)qB(x_1-x_{01})+\partial_2(q\varphi(x_i))\right),
\end{equation}
\textcolor{black}{for $\hat{M}_3$},
\begin{eqnarray}
\hat{M}_3:\ \ && -i\hbar\left[ (x_1-x_{01})\partial_2 - (x_2-x_{02})\partial_1\right]  \\
&& +\left[\frac{1}{2}\alpha\, qB\left((x_1-x_{01})^2-(x_2-x_{02})^2\right)-(x_1-x_{01})\partial_2(q\varphi(x_i))+(x_2-x_{02})\partial_1(q\varphi(x_i))\right], \nonumber
\end{eqnarray}
and for the velocity momenta \textcolor{black}{$\hat{p}_i$ ($i=1,2$)},
\begin{equation}
\hat{p}_1:\ \ -i\hbar\partial_1-\left(-\frac{1}{2}(\alpha+1)qB(x_2-x_{02})+\partial_1 (q\varphi(x_i))\right),
\end{equation}
\begin{equation}
\hat{p}_2:\ \ -i\hbar\partial_2-\left(-\frac{1}{2}(\alpha-1)qB(x_1-x_{01})+\partial_2(q\varphi(x_i))\right).
\end{equation}

In the discussion hereafter, the configuration space wave function representations of the bases which diagonalise the pairs $(\hat{T}_1,\hat{H})$
or $(\hat{M}_3,\hat{H})$ are constructed explicitly, namely for \textcolor{black}{abstract} states $|T_1,E_n\rangle$ or $|s\hbar\ell,E_n\rangle$ such that
\begin{eqnarray}
\hat{T}_1|T_1,E_n\rangle &=& T_1\,|T_1,E_n\rangle,\qquad
\hat{M}_3|s\hbar\ell,E_n\rangle = s \hbar \ell\,|s\hbar\ell,E_n\rangle, \nonumber \\
\hat{H}|T_1,E_n\rangle &=& E_n|T_1,E_n\rangle,\qquad
\hat{H}|s\hbar\ell,E_n\rangle = E_n |s\hbar\ell,E_n\rangle,
\end{eqnarray}
with the orthonormalisation conditions,
\begin{equation}
\langle T_1,E_{n_1}|T'_1,E_{n_2}\rangle=\delta(T_1-T'_1)\,\delta_{{n_1},{n_2}},\qquad
\langle s\hbar\ell_1E_{n_1}|s\hbar\ell_2,E_{n_2}\rangle = \delta_{\ell_1,\ell_2}\,\delta_{{n_1},{n_2}}.
\end{equation}
As is well known the energy spectrum of \textcolor{black}{the quantum Landau problem, organised into} infinitely degenerate Landau levels, is given by,
\begin{equation}
E_n=\hbar\omega_c\left(n+\frac{1}{2}\right),\qquad n=0,1,2,\cdots,
\end{equation}
with $T_1\in\mathbb{R}$ and $\ell\in\mathbb{Z}$, while $\ell\ge -n$ for any given value of $n=0,1,2,\cdots$.

The construction of the states $|T_2,E_n\rangle$ that diagonalise the pair of physical observables $(\hat{T}_2,\hat{H})$ is not included here.
It would proceed along lines similar to that of the states $|T_1,E_n\rangle$
(and \textcolor{black}{in fact} amounts \textcolor{black}{simply} to the exchange of coordinate axes $x_1\rightarrow x_2$ and $x_2\rightarrow -x_1$).

\subsection{The $|T_1,E_n\rangle$ basis}
\label{Sect5.1}

Based on the above representations of the \textcolor{black}{gauge invariant quantum} physical operators $\hat{T}_1$ and $\hat{H}$ in terms of
differential operators in $\partial_i$, it is a standard and direct exercise to establish that given all our choices of phase conventions
and canonical unitary \textcolor{black}{configuration space} representation of the Heisenberg algebra,
the configuration space wave functions of the states $|T_1,E_n\rangle$, with their normalisation as specified above,
are expressed as\footnote{Up to a global constant phase factor common to all these states, which is set to unity in this \textcolor{black}{final expression}.},
\begin{eqnarray}
&&\langle\vec{x}|T_1,E_n\rangle = e^{\frac{i}{\hbar}\left[\frac{1}{2}(1-\alpha)qB(x_1-x_{01})(x_2-x_{02})+q\varphi(x_i)\right]}\,
\frac{1}{\sqrt{2\pi\hbar}} e^{\frac{i}{\hbar}T_1(x_1-x_{01})} \\
&& \times \left(\frac{m\omega_c}{\pi\hbar}\right)^{1/4}\frac{1}{\sqrt{2^n\cdot n!}}
e^{-\frac{m\omega_c}{2\hbar}\left(x_2-\left(x_{02}-\frac{T_1}{qB}\right)\right)^2}
H_n\left(\sqrt{\frac{m\omega_c}{\hbar}}\left(x_2-\left(x_{02}-\frac{T_1}{qB}\right)\right)\right), \nonumber
\label{eq:Hn}
\end{eqnarray}
where $H_n(u)$ are the Hermite polynomials for $n=0,1,2,\cdots$.

Note well that the combination
\begin{equation}
x_{02}-\frac{T_1}{qB}=x_{c2},
\end{equation}
which contributes to the dependency in $x_2$ of these wave functions, corresponds precisely to the value of the second cartesian component of the
magnetic centre position vector. Indeed since as quantum operators they do not commute and define a Heisenberg algebra of their own,
the two components of that magnetic centre position vector cannot be diagonalised jointly. The states $|T_1,E_n\rangle$ are also
eigenstates of the gauge invariant physical observable $\hat{x}_{c2}$, with eigenvalue $x_{c2}=x_{02}-T_1/(qB)$,
namely $\hat{x}_{c2}|T_1,E_n\rangle=(x_{0i}-T_1/(qB))|T_1,E_n\rangle$. While they are maximally
delocalised relative to the coordinate $x_1$, with a uniform probability distribution along that direction in the plane.

Furthermore the presence of the overall phase factor in the above expression for $\langle \vec{x}\,|T_1,E_n\rangle$, \textcolor{black}{namely,}
\begin{equation}
e^{\frac{i}{\hbar}\left[\frac{1}{2}(1-\alpha)qB(x_1-x_{01})(x_2-x_{02})+q\varphi(x_i)\right]}=
e^{\frac{i}{\hbar}\left[\frac{1}{2}qB(x_1-x_{01})(x_2-x_{02})+q\bar{\varphi}(x_i)\right]},
\end{equation}
which includes the dependency on the gauge fixing parameters $(\alpha,\varphi(\vec{x}\,))$,
may be understood as follows. \textcolor{black}{In the absence of} that phase factor the above result is that obtained by working directly
in the Landau gauge with $(\alpha_2=1,\varphi_2(\vec{x}\,)=0)$, thus corresponding to a gauge transformation
away from the gauge $(\alpha,\varphi(\vec{x}\,))$ with given values \textcolor{black}{for} these parameters, generated by the function
\begin{equation}
\Delta\bar{\varphi}(\vec{x}\,)=-\frac{1}{2}(1-\alpha)B(x_1-x_{01})(x_2-x_{02})-\varphi(\vec{x}\,)=-\frac{1}{2}B(x_1-x_{01})(x_2-x_{02})-\bar{\varphi}(\vec{x}\,),
\end{equation}
in the notations of (\ref{eq:Delta},\ref{eq:psi}). Thus in that Landau gauge \textcolor{black}{with $(\alpha_2=1,\varphi_2(\vec{x}\,)=0)$}
one would have for any abstract \textcolor{black}{gauge invariant} quantum state,
\begin{equation}
\langle\vec{x};2|\psi\rangle=e^{\frac{i}{\hbar}q\Delta\bar{\varphi}(\vec{x}\,)}\,\langle\vec{x}\,|\psi\rangle,
\end{equation}
and in particular for the eigenstates $|T_1,E_n\rangle$ \textcolor{black}{in the $(\alpha,\varphi(\vec{x}\,))$ gauge},
\begin{equation}
\langle \vec{x}\,|T_1,E_n\rangle = e^{-\frac{i}{\hbar}q\Delta\bar{\varphi}(\vec{x}\,)}\,\langle \vec{x};2|T_1,E_n\rangle,
\end{equation}
with $\langle \vec{x};2|T_1,E_n\rangle$ thus given by the expression in (\ref{eq:Hn}) without the overall phase factor.
These features thus account for the necessary appearance of that overall phase factor for the wave functions $\langle \vec{x}\,|T_1,E_n\rangle$
in the general $(\alpha,\varphi(\vec{x}\,))$ gauge, by having chosen the canonical unitary \textcolor{black}{configuration space}
representation of the Heisenberg algebra irrespective of the chosen values for these gauge \textcolor{black}{fixing} parameters.

\subsection{The $|s\hbar\ell,E_n\rangle$ basis}
\label{Sect5.2}

Motivated by the observations made at the very end of Subection~\ref{Sect5.1}, in order to identify the wave functions $\langle \vec{x}\,|s\hbar\ell,E_n\rangle$
for the eigenstate basis of the commuting \textcolor{black}{gauge invariant}
operator pair $(\hat{M}_3,\hat{H})$ in the most efficient way possible\footnote{However, see also the comments
made in the very last two paragraphs of the present Subsection~\ref{Sect5.2}.},
let us apply to the gauge choice of specific \textcolor{black}{gauge fixing} parameters $(\alpha,\varphi(\vec{x}\,))$
the gauge transformation to the symmetric gauge with $(\alpha_2=0,\varphi_2(\vec{x}\,)=0)$, thus this time with
\begin{equation}
\Delta\bar{\varphi}(\vec{x}\,)=-\frac{1}{2} (-\alpha) B (x_1-x_{01})(x_2 - x_{02}) - \varphi(\vec{x}\,)=-\bar{\varphi}(\vec{x}\,),
\end{equation}
so that in this case one has $|\vec{x};2\rangle = e^{-\frac{i}{\hbar}q\Delta\bar{\varphi}(\vec{x}\,)}\,|\vec{x}\,\rangle$ as well as,
\begin{equation}
\psi(\vec{x}\,)=\langle \vec{x}\,|\psi\rangle= e^{-\frac{i}{\hbar}q\Delta\bar{\varphi}(\vec{x}\,)}\,\langle\vec{x};2|\psi\rangle
=e^{-\frac{i}{\hbar}q\Delta\bar{\varphi}(\vec{x}\,)}\,\psi_{(2)}(\vec{x}\,)=e^{\frac{i}{\hbar}q\bar{\varphi}(\vec{x}\,)}\,\psi_{(2)}(\vec{x}\,).
\end{equation}

By working now in terms of the configuration space wave function representation related to the basis $|\vec{x};2\rangle$ and the corresponding
wave functions $\psi_{(2)}(\vec{x}\,)=\langle \vec{x};2|\psi\rangle$, the relevant conserved and \textcolor{black}{gauge invariant physical} operators are realised
by the following differential operators,
\begin{equation}
\hat{H}:\ \ \ \frac{1}{2m}\left[-i\hbar\partial_1+\frac{1}{2}qB(x_2-x_{02})\right]^2+\frac{1}{2m}\left[-i\hbar\partial_2-\frac{1}{2}qB(x_1-x_{01})\right]^2,
\end{equation}
\begin{equation}
\hat{T}_1:\ \ \ -i\hbar\partial_1 - \frac{1}{2}qB(x_2-x_{02});\qquad\qquad
\hat{T}_2:\ \ \ -i\hbar\partial_2 + \frac{1}{2}qB(x_1-x_{01}),
\end{equation}
and,
\begin{equation}
\hat{M}_3:\ \ \ -i\hbar\epsilon_{ij}(x_i-x_{0i})\partial_j,
\end{equation}
while,
\begin{equation}
\hat{p}_1=\hat{T}_1+qB(\hat{x}_2-x_{02}\mathbb{I}),\qquad
\hat{p}_2=\hat{T}_2-qB(\hat{x}_1-x_{01}\mathbb{I}).
\end{equation}
The usual resolution in the symmetric gauge in terms of abstract Fock algebra generators may thus \textcolor{black}{readily} proceed in this case as well.

First introduce the following cartesian Fock generators,
\begin{equation}
a_i:\ \ \ \frac{1}{2}\sqrt{\frac{m\omega_c}{\hbar}}\left[\left(x_i-x_{0i}\right)+\frac{2\hbar}{m\omega_c}\partial_i\right],\qquad
a^\dagger_i:\ \ \ \frac{1}{2}\sqrt{\frac{m\omega_c}{\hbar}}\left[\left(x_i-x_{0i}\right)-\frac{2\hbar}{m\omega_c}\partial_i\right],
\end{equation}
such that the \textcolor{black}{abstract} phase space coordinate operators $(\hat{x}_i,\hat{p}_i)$ are expressed by,
\begin{equation}
\hat{x}_i-x_{0i}\mathbb{I}:\ \ \ \sqrt{\frac{\hbar}{m\omega_c}}\left(a_i + a^\dagger_i\right),\qquad
-i\hbar\partial_i:\ \ \ -\frac{i}{2}m\omega_c\sqrt{\frac{\hbar}{m\omega_c}}\left(a_i - a^\dagger_i\right),
\label{eq:xi}
\end{equation}
while $(a_i,a^\dagger_i)$ indeed generate a normalised Fock algebra,
\begin{equation}
\left[a_i,a^\dagger_j\right]=\delta_{ij}\,\mathbb{I}.
\end{equation}
Next introduce the normalised helicity Fock algebra generators, which diagonalise the canonical angular-momentum $\hat{L}^c_3[\vec{A}\,]$ in the symmetric gauge,
\begin{equation}
a_\pm=\frac{1}{\sqrt{2}}\left(a_1 \mp is a_2\right),\qquad
a^\dagger_\pm=\frac{1}{\sqrt{2}}\left(a^\dagger_1 \pm i s a^\dagger_2\right),\qquad
\left[a_\pm,a^\dagger_\pm\right]=\mathbb{I}.
\end{equation}
In terms of these linear operator redefinitions, one finds for the \textcolor{black}{abstract} conserved and \textcolor{black}{gauge invariant} physical Noether charges,
\begin{equation}
\hat{H}:\ \ \ \hbar\omega_c\left(a^\dagger_- a_- + \frac{1}{2}\right),
\end{equation}
\begin{equation}
\hat{T}_1:\ \ \ i\sqrt{\frac{\hbar m \omega_c}{2}}\left(a^\dagger_+ - a_+\right),\qquad\qquad
\hat{T}_2:\ \ \ s\sqrt{\frac{\hbar m \omega_c}{2}}\left(a^\dagger_+ + a_+\right),
\label{eq:T1-T2}
\end{equation}
and,
\begin{equation}
\hat{M}_3:\ \ \ s\hbar\left(a^\dagger_+ a_+ - a^\dagger_- a_-\right),
\end{equation}
while,
\begin{equation}
\hat{p}_1:\ \ \ i\sqrt{\frac{\hbar m\omega_c}{2}}\left(a^\dagger_- - a_-\right),\qquad\qquad
\hat{p}_2:\ \ \ -s\sqrt{\frac{\hbar m\omega_c}{2}}\left(a^\dagger_- + a_-\right).
\label{eq:p1-p2}
\end{equation}
Note how the two helicity Fock sectors of the quantised system correspond, on the one hand, to the magnetic centre operators $\hat{x}_{ci}$, or equivalently
the spatial translation generators $\hat{T}_i$, for the sector $(a_+,a^\dagger_+)$ which indeed maps quantum states only within a same Landau level,
and on the other hand, to the velocity momentum $\hat{p}_i$ which indeed necessarily maps between adjacent Landau levels separated by a single
quantum of energy $\hbar\omega_c$, these two ensembles of transformations indeed commuting
with one another\footnote{Each of the four possible pairs of operators $(\hat{T}_i,\hat{p}_j)$, or equivalently $(\hat{x}_{ci},\hat{p}_j)$,
thus provide yet other maximal ensembles of commuting gauge invariant and physical observables in terms of which to construct
other possible complete eigenbases of Hilbert space, then labelled by eigenvalues of one of the cartesian components of the magnetic centre
and one of the cartesian components of the velocity momentum. This observation of course agrees with the results already noted in (\ref{eq:triple}).}.

Consequently the joint diagonalisation of $(\hat{M}_3,\hat{H})$ is straightforward enough now in terms of the orthonormalised helicity Fock basis $|n_+,n_-\rangle$
$(n_+,n_-=0,1,2,\cdots)$ defined by
\begin{equation}
|n_+,n_-\rangle = \frac{1}{\sqrt{n_+!\,n_-!}}\,\left(a^\dagger_+\right)^{n_+}\,\left(a^\dagger_-\right)^{n_-}\,|\Omega\rangle,\qquad
\langle n_+,n_-|m_+,m_-\rangle=\delta_{n_+,m_+}\,\delta_{n_-,m_-},
\end{equation}
where the Fock vacuum $|\Omega\rangle$ is such that,
\begin{equation}
a_\pm|\Omega\rangle=0,\qquad
\langle\Omega|\Omega\rangle=1.
\end{equation}
For the configuration space wave function of that Fock vacuum, in which an arbitrary constant phase factor has been set to unity, one finds,
\begin{equation}
\langle\vec{x}; 2|\Omega\rangle=\left(\frac{m\omega_c}{2\pi\hbar}\right)^{1/2}\,e^{-\frac{m\omega_c}{4\hbar}(\vec{x}-\vec{x}_0)^2}.
\end{equation}
Applying then the general method outlined in \cite{Lee}, one finally finds explicitly for all Fock states, inclusive of a \textcolor{black}{necessary} phase factor $(-1)^n$,
\begin{equation}
\langle\vec{x};2|n_+,n_-\rangle=\left(\frac{m\omega_c}{2\pi\hbar}\right)^{1/2}\frac{(-1)^n\,n!}{\sqrt{n_+!\,n_-!}}\,e^{is\ell\theta}\,v^{|\ell|}\,e^{-\frac{1}{2}v^2}\,
L^{|\ell|}_n(v^2),
\end{equation}
where $L^m_n(u)$ are the Laguerre polynomials, while the quantities $n$ and $\ell$ are defined in terms of $(n_+,n_-)$ by,
\begin{equation}
n={\rm min}\,(n_+,n_-),\quad
\ell=n_+-n_-,\quad
{\rm max}\,(n_+,n_-)=n+|\ell|,\quad
n_\pm=n+\frac{|\ell|\pm \ell}{2},
\end{equation}
and the variables $u$, $\theta$ and $v$ by the changes of variables,
\begin{equation}
v=\sqrt{\frac{m\omega_c}{2\hbar}}\,u,\qquad
u=\sqrt{(\vec{x}-\vec{x}_0)^2},\qquad
x_1-x_{01}=u\cos\theta,\qquad
x_2-x_{02}=u\sin\theta,
\end{equation}
or in other words,
\begin{equation}
v=\sqrt{\frac{m\omega_c}{2\hbar}}|\vec{x}-\vec{x}_0|,\qquad
(x_1-x_{01})+i(x_2-x_{02})=u\,e^{i\theta}.
\end{equation}

In conclusion the orthonormalised basis $|s\hbar\ell,E_n\rangle$ of $(\hat{M}_3,\hat{H})$ joint eigenstates such that
\begin{equation}
\hat{M}_3|s\hbar\ell,E_n\rangle=s\hbar\ell\,|s\hbar\ell,E_n\rangle,\qquad
\hat{H}|s\hbar\ell,E_n\rangle = E_n\,|s\hbar\ell,E_n\rangle,
\end{equation}
is provided by the helicity Fock states $|n_+,n_-\rangle$ with the following correspondencies
between the values of $\ell$, $n$, and $n_\pm$,
\begin{equation}
|s\hbar\ell,E_n\rangle \equiv |n_+,n_-\rangle:\qquad
n_+=n+\ell,\qquad n_-=n.
\end{equation}
The spectrum of eigenvalues is thus organised in terms of the following ensemble of values for the integers $(n,\ell)$,
\begin{equation}
n=0,1,2,\cdots,\qquad \ell\ge -n,
\end{equation}
with the energy eigenvalues $E_n=\hbar\omega_c(n+1/2)$, thereby displaying the
infinite degeneracy in energy for the generalised angular-momentum values $(s\hbar\,\ell\ge -s\hbar\, n)$ within each of the Landau levels
in energy labelled by $n=0,1,2,\cdots$.

While finally, the configuration space wave functions of the eigenstates $|s\hbar\ell,E_n\rangle$ in the general $(\alpha,\varphi(\vec{x}\,))$ gauge choice are
provided by the expressions, inclusive of the \textcolor{black}{necessary} phase factors $(-1)^n$ and now $e^{\frac{i}{\hbar}q\bar{\varphi}(\vec{x}\,)}$ as well,
\begin{equation}
\langle \vec{x}\,|s\hbar\ell,E_n\rangle=e^{\frac{i}{\hbar}q\bar{\varphi}(x_i)}\,\left(\frac{m\omega_c}{2\pi\hbar}\right)^{1/2}(-1)^n\sqrt{\frac{n!}{(n+|\ell|)!}}\,
e^{is\ell\theta}\,v^{|\ell|}\,e^{-\frac{1}{2}v^2}\,L^{|\ell|}_n(v^2),
\end{equation}
with the change of variables $(x_1,x_2)\leftrightarrow (v,\theta)$ defined by
\begin{equation}
(x_1-x_{01})+i(x_2-x_{02})=\sqrt{\frac{2\hbar}{m\omega_c}}\,v\,e^{i\theta}.
\end{equation}

Concurrently with the last result, let us emphasize the following relevant and general point. Given any abstract quantum operator $\hat{\cal O}$
which in the wave function representation $\langle\vec{x};2|\psi\rangle$ for any abstract \textcolor{black}{gauge invariant} quantum state $|\psi\rangle$
in the $(\alpha_2=0,\varphi_2(\vec{x}\,)=0)$ symmetric gauge is represented by the differential operator ${\cal O}_{(0,0)}(x_i,\partial_i)$ as,
\begin{equation}
\langle \vec{x};2|\hat{\cal O}|\psi\rangle = {\cal O}_{(0,0)}(x_i,\partial_i)\,\langle\vec{x};2|\psi\rangle,
\end{equation}
then in the general gauge of parameters $(\alpha,\varphi(\vec{x}\,))$ the same \textcolor{black}{abstract}
operator $\hat{\cal O}$ is represented by the differential operator
\begin{equation}
{\cal O}_{(\alpha,\varphi)}(x_i,\partial_i)=e^{\frac{i}{\hbar}q\bar{\varphi}(\vec{x}\,)}\,{\cal O}_{(0,0)}(x_i,\partial_i)\,e^{-\frac{i}{\hbar}q\bar{\varphi}(\vec{x}\,)},
\end{equation}
namely such that,
\begin{equation}
\langle \vec{x}\,|\hat{\cal O}|\psi\rangle = {\cal O}_{(\alpha,\varphi)}(x_i,\partial_i)\,\langle \vec{x}\,|\psi\rangle.
\end{equation}
In other words, through this redefinition of the differential operator representations of abstract quantum operators, the above (cartesian and helicity) Fock algebras
with all their abstract calculational advantages that avoid the explicit resolution of differential equations and overlap integrals, may be introduced
and exploited in precisely the same manner as well given any other arbitrary gauge choice $(\alpha,\varphi(\vec{x}\,))$,
and not only for the symmetric gauge with $(\alpha_2=0,\varphi_2(\vec{x}\,)=0)$.
Quite obviously the same general comment extends to the differential operator representations of a same abstract quantum operator, but considered
for different general gauge choices $(\alpha_1,\varphi_1(\vec{x}\,))$ and $(\alpha_2,\varphi_2(\vec{x}\,))$, by accounting for the necessary phase factor
$e^{\frac{i}{\hbar}\Delta\bar{\varphi}(\vec{x}\,)}$ in terms of the gauge transformation function $\Delta\bar{\varphi}(\vec{x}\,)$ considered
in (\ref{eq:Delta},\ref{eq:psi}), once a same specific unitary \textcolor{black}{configuration space}
representation of the Heisenberg algebra of the abstract \textcolor{black}{canonical} conjugate momenta $\hat{\pi}[\vec{A}\,]_i$
is chosen irrespective of the values for the gauge \textcolor{black}{fixing} parameters $(\alpha,\varphi(\vec{x}\,))$.

All possible unitary \textcolor{black}{configuration space} representations of the quantum Landau problem,
for all possible choices of unitary \textcolor{black}{configuration space}
representations of the Heisenberg algebra of its \textcolor{black}{abstract} conjugate momenta $\hat{\pi}[\vec{A}\,]_i$,
and for any possible choice of gauge for its vector gauge potential
$\vec{A}(\vec{x}\,)$, are thus certainly all unitarily equivalent and all physically identical for all its gauge invariant and thus physical quantum observables.
All possible canonical quantisations of the Landau problem for whatever choice of gauge for its vector potential belong to a same and
unique U(1) gauge equivalence class.

\subsection{The change of basis $\langle n_+,n_- | T_1,E_{m_-}\rangle$}
\label{Sect5.3}

In view of this unitary equivalence of all possible realisations of the quantum Landau problem,
and having identified the configuration space wave functions of the joint \textcolor{black}{abstract and gauge invariant} eigenstates $|T_1,E_n\rangle$ and
$|s\hbar\ell,E_n\rangle=|n_+=n+\ell,n_-=n\rangle$ $(n\in\mathbb{N}, \ell\ge n)$ for the pairs
of conserved \textcolor{black}{gauge invariant} observables $(\hat{T}_1,\hat{H})$ and $(\hat{M}_3,\hat{H})$, respectively, for a general $(\alpha,\varphi(\vec{x}\,))$
gauge choice for the background vector gauge field representing the applied background static and uniform magnetic field $\vec{B}$,
it is now be possible to determine the matrix elements for the change of basis between these two choices of joint eigenstates, or equivalently the
\textcolor{black}{gauge invariant} matrix elements $\langle n_+,n_-|T_1,E_{m_-}\rangle$. These matrix elements are thus given by,
\begin{equation}
\langle n_+,n_-|T_1,E_{m_-}\rangle = \int_{\mathbb{R}^2} d^2\vec{x}\, \langle n_+,n_-|\vec{x}\,\rangle\,\langle \vec{x}\,|T_1,E_{m_-}\rangle,
\end{equation}
in terms of the wave functions $\langle\vec{x}\,|T_1,E_{m_-}\rangle$ and
$\langle \vec{x}\,|n_+,n_-\rangle=e^{\frac{i}{\hbar}q\bar{\varphi}(\vec{x}\,)}\langle\vec{x};2|n_+,n_-\rangle$ constructed above.
Incidentally note that necessarily, these matrix elements are diagonal in $(n_-,m_-)$ since the two helicity Fock sectors commute with one another,
so that $\langle n_+,n_-|T_1,E_{m_-}\rangle=\langle n_+|T_1\rangle\,\delta_{n_-,m_-}$, where the matrix elements $\langle n_+|T_1\rangle$ remain to be computed
in the helicity Fock sector $(a_+,a^\dagger_+)$ (indeed, $\hat{T}_1$ and $\hat{T}_2$ only act in that sector of Hilbert space: see (\ref{eq:T1-T2}).
While $(\hat{p}_1,\hat{p}_2)$ only act in the complementary helicity Fock sector $(a_-,a^\dagger_-)$: see (\ref{eq:p1-p2})).

However rather than proceed with a direct but indeed tedious evaluation of these overlapping integrals \textcolor{black}{in terms of
configuration space wave functions} as is done in \cite{Haugset,Waka2} through a series
of changes of variables using a specific integral representation of Laguerre polynomials in terms of a product of two Hermite polynomials,
here let us take advantage of the relations in (\ref{eq:T1-T2}) between the Noether translation generators $\hat{T}_i$ and the helicity Fock algebra
$(a_+,a^\dagger_+)$, which translate into,
\begin{equation}
a_+=\frac{i}{\sqrt{2\hbar|qB|}}\left(\hat{T}_1 - i s \hat{T}_2\right),\qquad
a^\dagger_+=\frac{-i}{\sqrt{2\hbar|qB|}}\left(\hat{T}_1 + i s \hat{T}_2\right).
\end{equation}
Given this observation, the matrix elements $\langle T_1|n_+\rangle$ are nothing else, up to some different physical normalisation and phase factors,
than the usual configuration space wave functions of the Fock states of \textcolor{black}{the ordinary} one dimensional harmonic oscillator
of coordinate $T_1$ and conjugate momentum $T_2$.

The result for $\langle n_+,n_-|T_1,E_{m_-}\rangle$
then readily follows after some simple algebra, up to some constant phase factor that remains to be identified \textcolor{black}{but in fact} corresponds
to the phase factor of \textcolor{black}{the quantity} $\langle n_+=0|T_1\rangle$ which \textcolor{black}{itself} is normalised such that
$\int _{-\infty}^{+\infty} dT_1 |\langle T_1|n_+=0\rangle|^2=1$.
To determine the value of that phase factor given all the previous choices
of phases for the wave functions $\langle\vec{x}\,|n_+,n_-\rangle$ and $\langle \vec{x}\,|T_1,E_{n_-}\rangle$, one need only evaluate explicitly using these
wave functions, say the matrix element $\langle n_+=0,n_-=0|T_1,E_{(m_-=0)}\rangle$, which reduces to an exercise in gaussian integrals in $\vec{x}$.

Doing so, one finds that the matrix elements for the relevant change of basis are explicitly given by,
\begin{equation}
\langle n_+,n_-|T_1,E_{m_-}\rangle = \frac{i^{n_+}}{\sqrt{2^{n_+}\cdot n_+!}}\left(\pi\hbar m\omega_c\right)^{-1/4}\,e^{-\frac{1}{2}\frac{T^2_1}{\hbar m\omega_c}}\,
H_{n_+}\left(\frac{T_1}{\sqrt{\hbar m\omega_c}}\right)\,\delta_{n_-,m_-},
\label{eq:change}
\end{equation}
a result which \textcolor{black}{concurs with expressions given in Refs.\cite{Haugset,Waka2} (which however, do not specifiy their phase conventions)
as far as may be checked.} Note how the change of basis matrix elements
$\langle n_+,n_-|T_1,E_{m_-}\rangle$ are diagonal in $(n_-,m_-)$\textcolor{black}{, as it should of course}.

Incidentally, from the integral representation
\begin{eqnarray}
\langle\vec{x}\,|n_+,n_-\rangle &=& \sum_{m_-=0}^\infty\int_{-\infty}^{+\infty}dT_1\,\langle\vec{x}\,|T_1,E_{m_-}\rangle\langle T_1,E_{m_-}|n_+,n_-\rangle \nonumber \\
&=& \int ^{+\infty}_{-\infty}dT_1\,\langle\vec{x}\,|T_1,E_{n_-}\rangle \langle T_1,E_{n_-}|n_+,n_-\rangle,
\end{eqnarray}
\textcolor{black}{and} by using the above result for $\langle n_+,n_-|T_1,E_{n_-}\rangle$ one obtains an integral representation of Laguerre polynomials
in terms of the product of two Hermite polynomials with shifted arguments and of different orders\textcolor{black}{, which is not provided explicitly here}.

On the other hand, one also has,
\begin{equation}
\int_{-\infty}^{+\infty}dT_1\,|T_1,E_{n_-}\rangle\,\frac{(-i)^{n_+}}{\sqrt{2^{n_+}\cdot n_+!}}\left(\pi\hbar m\omega_c\right)^{-1/4}
\,e^{-\frac{1}{2}\frac{T^2_1}{\hbar m\omega_c}}\,H_{n_+}\left(\frac{T_1}{\sqrt{\hbar m \omega_c}}\right) = |n_+,n_- \rangle ,
\end{equation}
simply because \textcolor{black}{of the completeness relation for the states $|T_1,E_{m_-}\rangle$},
\begin{equation}
|n_+,n_-\rangle=\sum_{m_-=0}^\infty\int_{-\infty}^{+\infty} dT_1\,|T_1,E_{m_-}\rangle\,\langle T_1,E_{m_-}|n_+,n_-\rangle .
\label{eq:identity}
\end{equation}

\subsection{A sample of matrix elements}
\label{Sect5.4}

In view of the relations (\ref{eq:Relation1},\ref{eq:Relation2},\ref{eq:Relation3}), the matrix elements for any pair of abstract quantum states $|\psi_1\rangle$
and $|\psi_2\rangle$ of the \textcolor{black}{abstract} canonical conjugate momenta $\hat{\pi}[\vec{A}\,]_i$ and canonical angular-momentum
$\hat{L}^c_3[\vec{A}\,]$ operators\textcolor{black}{---which are not gauge invariant\footnote{\textcolor{black}{Given the gauge transformed vector potential
$\vec{A}_2(\vec{x}\,)=\vec{A}(\vec{x}\,)+\vec{\nabla}\Delta\bar{\varphi}(\vec{x}\,)$, these operators transform as
$\hat{\pi}[\vec{A}_2]_i=\hat{\pi}[\vec{A}\,]_i+\partial_i\Delta\bar{\varphi}(\hat{\vec{x}})$ as well as
$\hat{L}^c_3[\vec{A}_2]=\hat{L}^c_3[\vec{A}\,]+\epsilon_{ij}(\hat{x}_i-x_{0i}\mathbb{I})\partial_j\Delta\bar{\varphi}(\hat{\vec{x}})$.}}---}are 
given in terms of matrix elements of the gauge invariant and physical Noether charges $\hat{T}_i$ and $\hat{M}_3$ as follows,
\begin{eqnarray}
\langle\psi_1|\hat{\pi}[\vec{A}\,]_i|\psi_2\rangle  &=& \langle\psi_1|\hat{T}_i|\psi_2\rangle
+ \langle\psi_1|\frac{1}{2}qB\epsilon_{ij}(\hat{x}_j-x_{0j}\mathbb{I})|\psi_2\rangle +\langle\psi_1|\partial_i(q\bar{\varphi}(\hat{x}_i))|\psi_2\rangle, \nonumber \\
\langle\psi_1|\hat{L}^c_3[\vec{A}\,]|\psi_2\rangle &=& \langle\psi_1|\hat{M}_3|\psi_2\rangle +
\langle\psi_1|\epsilon_{ij}(\hat{x}_i-x_{0i}\mathbb{I})\partial_j(q\bar{\varphi}(\hat{x}_i))|\psi_2\rangle \nonumber \\
&=& \langle\psi_1|\hat{M}_3|\psi_2\rangle  -\frac{1}{2}\alpha qB\langle\psi_1|\left[(\hat{x}_1-x_{01}\mathbb{I})^2-(\hat{x}_2-x_{02}\mathbb{I})^2\right]|\psi_2\rangle
\nonumber \\
&& + \langle\psi_1|\left[(\hat{x}_1-x_{01}\mathbb{I})\partial_2(q\varphi(\hat{x}_i) - (\hat{x}_2-x_{02}\mathbb{I})\partial_1(q\varphi(\hat{x}_i)\right]|\psi_2\rangle.
\end{eqnarray}
\textcolor{black}{These relations clearly display} that the canonical quantum operators $\hat{\pi}[\vec{A}\,]_i$ and $\hat{L}^c_3[\vec{A}\,]$
and \textcolor{black}{therefore} any of their matrix elements
are \textcolor{black}{indeed} certainly not gauge invariant, and thus certainly not physical either. \textcolor{black}{In particular the values of their matrix elements
are explicitly function of the gauge fixing parameters $(\alpha,\varphi(\vec{x}\,))$}. Nonetheless and as was noted previously,
in the symmetric gauge $(\alpha=0,\varphi(\vec{x}\,)=0)$
and thus $\bar{\varphi}(\vec{x}\,)=0$ the canonical angular-momentum operator $\hat{L}^c_3[\vec{A}\,]$ \textcolor{black}{hence all its matrix elements
as well coincide with} the Noether charge operator $\hat{M}_3$ as the generator of spatial rotations about $\vec{x}_0$ \textcolor{black}{and all its same matrix
elements}. Likewise for the conjugate momenta operators \textcolor{black}{$\hat{\pi}[\vec{A}\,]_i$},
since one has
\begin{eqnarray}
\langle\psi_1|\hat{\pi}[\vec{A}\,]_1|\psi_2\rangle &=& \langle\psi_1|\hat{T}_1|\psi_2\rangle
-\frac{1}{2}(\alpha-1)qB\langle\psi_1|(\hat{x}_2-x_{02}\mathbb{I})|\psi_2\rangle +\langle\psi_1|\partial_1(q\varphi(\hat{x}_i))|\psi_2\rangle, \nonumber \\
\langle\psi_1|\hat{\pi}[\vec{A}\,]_2|\psi_2\rangle &=& \langle\psi_1|\hat{T}_2|\psi_2\rangle
-\frac{1}{2}(\alpha+1)qB\langle\psi_1|(\hat{x}_1-x_{01}\mathbb{I})|\psi_2\rangle +\langle\psi_1|\partial_2(q\varphi(\hat{x}_i))|\psi_2\rangle,
\end{eqnarray}
 in the Landau gauge $(\alpha=1,\varphi(\vec{x}\,)=0)$ (resp., $(\alpha=-1,\varphi(\vec{x}\,)=0)$) the \textcolor{black}{canonical}
 conjugate momentum operator $\hat{\pi}[\vec{A}\,]_1$
 (resp., $\hat{\pi}[\vec{A}\,]_2$) coincides with the Noether charge operator $\hat{T}_1$ (resp., $\hat{T}_2$)
as the generator of spatial translations along the cartesian coordinate $x_1$ (resp., $x_2$).

For the evaluation of matrix elements of the Noether charges now specifically in the $|T_1,E_n\rangle$ and $|s\hbar\ell,E_n\rangle$ eigenbases,
in as much as feasible one may exploit the relationships that structure these matrix elements which derive from the algebra of commutation relations
of the quantum operators $\hat{H}$, $\hat{T}_i$, $\hat{M}_3$, $\hat{p}_i$ and $\hat{L}_3$, rather than through direct integral evaluations using
the wave function representations of these states \textcolor{black}{which were} established above.

For the operator $\hat{T}_1$ one obviously has,
\begin{equation}
\langle T_1,E_{n_1}|\hat{T}_1|T'_1,E_{n_2}\rangle=T_1\,\delta(T_1 - T'_1)\,\delta_{n_1,n_2}.
\end{equation}
The evaluation of the same matrix elements for the commutator $\left[\hat{T}_1,\hat{T}_2\right]=-is\hbar m \omega_c\mathbb{I}$ then implies the equation,
\begin{equation}
\left(T_1 - T'_1\right)\,\langle T_1,E_{n_1}|\hat{T}_2|T'_1,E_{n_2}\rangle=-i s \hbar m \omega_c\,\delta(T_1-T'_1)\,\delta_{n_1,n_2},
\end{equation}
of which the solution is
\begin{equation}
\langle T_1,E_{n_1}|\hat{T}_2|T'_1,E_{n_2}\rangle=\left(i\,s\hbar m \omega_c\,\delta'(T_1-T'_1) + A_{n_1}(T_1)\,\delta(T_1 - T'_1)\right)\,\delta_{n_1,n_2},
\end{equation}
where $A_{n_1}(T_1)$ is some specific function to be identified. Incidentally note how these matrix elements define a wave function representation in $T_1$
configuration space of the Heisenberg algebra spanned by $(\hat{T}_1,\hat{T}_2)$, with a unitary representation associated to a flat U(1) gauge connection
specified by $A_{n_1}(T_1)$. The determination of that quantity proceeds \textcolor{black}{from} a substitution of the above solution into the l.h.s.~of the
following identity based on (\ref{eq:identity}),
\begin{equation}
\int_{-\infty}^{+\infty}dT'_1\,\langle T_1,E_{n_1}|\hat{T}_2|T'_1,E_{n_1}\rangle\,\langle T'_1,E_{n_1}|n_+,n_-=n_1\rangle =
\langle T_1,E_{n_1}|\hat{T}_2|n_+,n_-=n_1\rangle.
\end{equation}
The r.h.s~ of this relation is readily evaluated using the expression (\ref{eq:T1-T2}) for $\hat{T}_2$ in terms of the helicity Fock algebra $(a_+,a^\dagger_+)$
(to be considered now within the general $(\alpha,\varphi(\vec{x}\,))$ gauge),
\begin{eqnarray}
&& \langle T_1,E_{n_1}|\hat{T}_2|n_+,n_-=n_1\rangle  \\
&& = s\sqrt{\frac{\hbar m \omega_c}{2}}
\left[\sqrt{n_++1}\,\langle T_1,E_{n_1}|n_+ + 1 , n_-=n_1\rangle + \sqrt{n_+}\,\langle T_1,E_{n_1}|n_+ - 1, n_-=n_1\rangle\right]. \nonumber
\end{eqnarray}
Using then the expression (\ref{eq:change}) for $\langle n_+,n_-|T_1,E_n\rangle$
and properties of the Hermite polynomials (namely, their three term recurrence relation \textcolor{black}{$H_{n+1}(x)-2xH_n(x)+2nH_{n-1}(x)=0$}
as well as the fact that $dH_n(x)/dx=2n H_{n-1}(x)$),
it easily follows that one must have $A_{n_1}(T_1)=0$. The matrix elements of both Noether charges $\hat{T}_i$ in the $|T_1,E_n\rangle$ basis
are thereby all determined, as listed in Table~\ref{Table1}. While their values in the $|s\hbar\ell,E_n\rangle$ basis as listed in Table~\ref{Table2}
are straightforwardly established using the helicity Fock algebra representations of these two operators, see (\ref{eq:T1-T2}).

The evaluation in the basis $|T_1,E_n\rangle$ of the commutation relation $\left[\hat{T}_1,\hat{M}_3\right]=-i\hbar \hat{T}_2$ implies the equation
\begin{equation}
\left(T_1 - T'_1\right)\langle T_1,E_{n_1}|\hat{M}_3|T'_1,E_{n_2}\rangle = -i\hbar\,\langle T_1,E_{n_1}|\hat{T}_2|T'_1,E_{n_2}\rangle
=s\,\hbar^2 m \omega_c\,\delta'(T_1-T'_1)\,\delta_{n_1,n_2},
\end{equation}
of which the solution is,
\begin{equation}
\langle T_1,E_{n_1}|\hat{M}_3|T'_1,E_{n_2}\rangle =
\left( -\frac{1}{2}s\,\hbar^2 m \omega_c\,\delta''(T_1 - T'_1) + B_{n_1}(T_1)\,\delta(T_1-T'_1)\right)\,\delta_{n_1,n_2},
\end{equation}
where $B_{n_1}(T_1)$ is some specific function to be identified.

Proceeding as above in the case of the matrix elements $\langle T_1,E_{n_1}|\hat{T}_2|T'_1,E_{n_1}\rangle$, by considering now the relation
\begin{eqnarray}
&&\int_{-\infty}^{+\infty}dT'_1\,\langle T_1,E_{n_1}|\hat{M}_3|T'_1,E_{n_1}\rangle\,\langle T'_1,E_{n_1}|n_+,n_-=n_1\rangle \nonumber \\
&=& \langle T_1,E_{n_1}|\hat{M}_3|n_+,n_-=n_1\rangle = s\hbar(n_+ - n_1)\,\langle T_1,E_{n_1}|n_+,n_-=n_1\rangle,
\end{eqnarray}
and using the differential equation that Hermite polynomials satisfy, namely $H''_n(x)-2 x H'_n(x)+ 2n H_n(x)=0$, it readily follows that
\begin{equation}
B_{n_1}(T_1)=s\hbar\left(\frac{T^2_1}{2\hbar m \omega_c}-(n_1+\frac{1}{2})\right),
\end{equation}
thereby providing the matrix elements $\langle T_1,E_{n_1}|\hat{M}_3|T'_1,E_{n_2}\rangle$ as listed in Table~\ref{Table1}.
While the obvious values for $\langle s\hbar\ell_1,E_{n_1}|\hat{M}_3|s\hbar\ell_2, E_{n_2}\rangle$ are listed in Table~\ref{Table2}.

\begin{table}
\centering
\begin{tabular}{| c || c |}
\hline
Operator $\hat{\cal O}$ & $\langle T_1,E_{n_1}|\hat{\cal O}| T'_1,E_{n_2}\rangle$  \\
\hline\hline
$\hat{T}_1$ & $T_1\,\delta(T_1 - T'_1)\,\delta_{n_1,n_2}$ \\
\hline
$\hat{T}_2$ & $ i\,s\,\hbar m \omega_c\,\delta'(T_1-T'_1)\,\delta_{n_1,n_2} $\\
\hline
$\hat{M}_3$ & $\left[-\frac{1}{2}s\,\hbar^2 m \omega_c\,\delta''(T_1 - T'_1) 
+ s\hbar\left(\frac{T^2_1}{2\hbar m \omega_c}-(n_1+\frac{1}{2})\right)\,\delta(T_1-T'_1) \right]\,\delta_{n_1,n_2}$  \\
\hline
$\hat{p}_1$ & $s\sqrt{\frac{\hbar m \omega_c}{2}}\left(\sqrt{n_1}\,\delta_{n_1,n_2+1} + \sqrt{n_2}\,\delta_{n_2,n_1+1}\right)\delta(T_1-T'_1)$  \\
\hline
$\hat{p}_2$ & $i\sqrt{\frac{\hbar m \omega_c}{2}}\left(\sqrt{n_1}\,\delta_{n_1,n_2+1} - \sqrt{n_2}\,\delta_{n_2,n_1+1}\right)\delta(T_1-T'_1)$  \\
\hline
$\hat{L}_3$ & For $n_1=n_2$:\ \ \ $ - s\hbar (2n_1+1)\,\delta(T_1 - T'_1)$ \\
\hline
\end{tabular}
\caption{\label{Table1} Matrix elements in the $|T_1,E_n\rangle$ eigenbasis.}
\end{table}

\begin{table}
\centering
\begin{tabular}{| c || c | c |}
\hline
Operator $\hat{\cal O}$ & $\langle s\hbar\ell_1,E_{n_1} | \hat{\cal O} | s\hbar \ell_2,E_{n_2}\rangle$ \\
\hline\hline
$\hat{T}_1$ & $i\sqrt{\frac{\hbar m\omega_c}{2}}\left(\sqrt{n_1+\ell_1}\,\delta_{\ell_1,\ell_2+1} - \sqrt{n_1+\ell_2}\,\delta_{\ell_2,\ell_1+1}\right)\,\delta_{n_1,n_2} $  \\
\hline
$\hat{T}_2$ &  $s\sqrt{\frac{\hbar m\omega_c}{2}}\left(\sqrt{n_1+\ell_1}\,\delta_{\ell_1,\ell_2+1} + \sqrt{n_1+\ell_2}\,\delta_{\ell_2,\ell_1+1}\right)\,\delta_{n_1,n_2} $\\
\hline
$\hat{M}_3$ & $s\hbar\ell_1\,\delta_{\ell_1,\ell_2}\,\delta_{n_1,n_2}$ \\
\hline
$\hat{p}_1$ & $i\sqrt{\frac{\hbar m \omega_c}{2}}\left(\sqrt{n_1}\,\delta_{\ell_2,\ell_1+1}\,\delta_{n_1,n_2+1} -
\sqrt{n_2}\,\delta_{\ell_1,\ell_2+1}\,\delta_{n_2,n_1+1}\right) $ \\
\hline
$\hat{p}_2$ & $-s \sqrt{\frac{\hbar m \omega_c}{2}}\left(\sqrt{n_1}\,\delta_{\ell_2,\ell_1+1}\,\delta_{n_1,n_2+1} +
\sqrt{n_2}\,\delta_{\ell_1,\ell_2+1}\,\delta_{n_2,n_1+1}\right) $ \\
\hline
$\hat{L}_3$ & For $n_1=n_2$:\ \ \ $ - s\hbar (2n_1+1)\,\delta_{\ell_1,\ell_2}$ \\
\hline
\end{tabular}
\caption{\label{Table2} Matrix elements in the $|s\hbar\ell,E_n\rangle$ eigenbasis.}
\end{table}

Since they commute with the quantum Hamiltonian $\hat{H}$, the gauge invariant and physical observables that are the Noether charges $\hat{T}_i$
and $\hat{M}_3$ and their matrix elements are of course diagonal in the energy eigenvalues. These three operators map between quantum states that
belong to a same Landau level, generally by changing their probability distribution in the $T_1$ variable or magnetic centre coordinate $x_{c2}$,
or in the total angular-momentum distribution in $s\hbar\ell$. In contradistinction the other three gauge invariant and physical observables of interest,
namely the velocity momenta $\hat{p}_i$ and the orbital angular-momentum $\hat{L}_3$, since they do not commute with the quantum Hamiltonian $\hat{H}$,
necessarily may map as well between different Landau levels, and generally also between different $T_1$ or $s\hbar\ell$ eigenvalues. Evaluating their matrix
elements is also of relevance.

Consider any pair of abstract quantum states of definite eigen-energies $E_{n_1}$ and $E_{n_2}$,
\textcolor{black}{namely} $|\lambda_1,E_{n_1}\rangle$ and $|\lambda_2,E_{n_2}\rangle$, $\lambda_{1,2}$
being additional labels to distinguish these states (which could stand for either the $T_1$ or the $s\hbar\ell$ eigenvalues, for instance). Evaluating for these states
the matrix elements of the commutation relations $\left[\hat{p}_i,\hat{H}\right]=i\,s\hbar\omega_c\epsilon_{ij}\,\hat{p}_j$ leads to a coupled system of two
linear equations for the matrix elements of $\hat{p}_i$, of which the solution is such that,
\begin{eqnarray}
&& \cdot\ {\rm If}\ \ (n_1-n_2)^2\ne 1:\quad
\langle \lambda_1,E_{n_1}|\hat{p}_1|\lambda_2,E_{n_2}\rangle=0=\langle\lambda_1, E_{n_1}|\hat{p}_2| \lambda_2,E_{n_2}\rangle , \nonumber \\
&& \cdot\ {\rm If}\ \ (n_1-n_2)^2=1:\quad
\langle\lambda_1,E_{n_1}|\hat{p}_2|\lambda_2,E_{n_2}\rangle = is(n_1 - n_2)\,\langle\lambda_1,E_{n_1}|\hat{p}_1|\lambda_2,E_{n_2}\rangle.
\end{eqnarray}
Using the explicit wave function representation for $\langle \vec{x}\,|T_1,E_n\rangle$ and the differential operator representation for the operator $\hat{p}_1$,
a simple calculation using the orthogonality properties of Hermite polynomials as well as the fact that $H'_n(x)=2n H_{n-1}(x)$ then finds,
\begin{equation}
\langle T_1,E_{n_1}|\hat{p}_1|T'_1,E_{n_2}\rangle = s\sqrt{\frac{\hbar m \omega_c}{2}}
\left[\sqrt{n_1}\,\delta_{n_1,n_2+1}\,+\,\sqrt{n_2}\,\delta_{n_2,n_1+1}\right]\,\delta(T_1 - T'_1).
\end{equation}
Consequently the same matrix elements for $\hat{p}_2$ may also be determined from the above relations, with final values listed in Table~\ref{Table1} as well.
While the matrix elements of these two operators in the $|s\hbar\ell,E_n\rangle$ eigenbasis readily follow once again from the representations of the
velocity momentum components $\hat{p}_i$ in terms of the helicity Fock algebra $(a_-,a^\dagger_-)$ (see (\ref{eq:p1-p2})), with the final values listed
in Table~\ref{Table2}. Note specifically that, as pointed out above already, the velocity momentum components $\hat{p}_i$ necessarily map between
Landau levels separated by a single quantum of the cyclotron frequency energy gap $\hbar\omega_c$. Thus in particular all their matrix elements for
external states of same energy, $n_1=n_2$, necessarily vanish \cite{Waka1}. Nevertheless these quantum operators do not vanish on Hilbert space.
The energy carried by the quantum velocity momenta $\hat{p}_i$ maps between adjacent Landau levels with the energy gap $\hbar\omega_c$.

Note that similar considerations for the matrix elements in the $|s\hbar\ell,E_n\rangle$ eigenbasis of the commutation relations
$\left[\hat{T}_i,\hat{M}_3\right]=-i\hbar\,\epsilon_{ij}\,\hat{T}_j$ lead to the following properties,
\begin{eqnarray}
&& \!\! \cdot\ {\rm If}\ \ (\ell_1-\ell_2)^2\ne 1:\ \ 
\langle s\hbar\ell_1,E_{n_1}|\hat{T}_1| s\hbar\ell_2,E_{n_2}\rangle=0=\langle s\hbar\ell_1, E_{n_1}|\hat{T}_2| s\hbar\ell_2,E_{n_2}\rangle , \nonumber \\
&&\!\! \cdot\ {\rm If}\ \ (\ell_1-\ell_2)^2=1:\ \
\langle s\hbar\ell_1,E_{n_1}|\hat{T}_2| s\hbar\ell_2,E_{n_2}\rangle = -is(\ell_1 - \ell_2)\,\langle\ell_1,E_{n_1}|\hat{T}_1| s\hbar\ell_2,E_{n_2}\rangle.
\end{eqnarray}
Of course, these conditions are met by the values for these matrix elements already listed in Table~\ref{Table2}.

In a similar manner, the commutator $\left[\hat{L}_3,\hat{M}_3\right]=0$ implies for the operator $\hat{L}_3$,
\begin{equation}
\langle s\hbar\ell_1,E_{n_1}|\hat{L}_3| s\hbar\ell_2,E_{n_2}\rangle=\langle s\hbar\ell_1,E_{n_1}|\hat{L}_3| s\hbar\ell_1, E_{n_2}\rangle\,\delta_{\ell_1,\ell_2}.
\end{equation}
In other words, the orbital angular-momentum operator $\hat{L}_3$, even when mapping between different Landau levels,
maps between quantum states with the same total angular-momentum distribution in $s\hbar\ell$.
On the other hand, given the commutation relation
\begin{equation}
\left[\hat{L}_3,\hat{H}\right]=-\frac{1}{2}i\,s \hbar\omega_c\left[(\hat{x}_i-x_{0i}\mathbb{I})\hat{p}_i+\hat{p}_i(\hat{x}_i-x_{0i}\mathbb{I})\right],
\end{equation}
one has,
\begin{equation}
(n_2-n_1)\langle s\hbar\ell_1,E_{n_1}|\hat{L}_3| s\hbar\ell_1,E_{n_2}\rangle\rangle = 
-i\,s\,\langle s\hbar\ell_1,E_{n_1} | \left[\frac{(\hat{x}_i-x_{0i}\mathbb{I})\hat{p}_i+\hat{p}_i(\hat{x}_i-x_{0i}\mathbb{I})}{2}\right] | s\hbar\ell_1,E_{n_2}\rangle.
\end{equation}
Thus an evaluation of the matrix elements of $\hat{L}_3$ must consider its explicit representation, say in terms of the helicity Fock algebra generators.
Using the relation $\hat{L}_3=\hat{M}_3-\frac{1}{2}qB\left(\hat{\vec{x}}-\vec{x}_0\mathbb{I}\right)^2$ as well as the expressions in (\ref{eq:xi}), one finds,
\begin{equation}
\hat{L}_3:\ \ \  -s \hbar\left(2 a^\dagger_- a_- + \mathbb{I} +(a^\dagger_+ a^\dagger_- + a_+ a _-)\right).
\label{eq:L3}
\end{equation}
This expression makes it explicit how through the contributions to it of the operators $(a_-,a^\dagger_-)$,
$\hat{L}_3$ maps between different Landau levels (with as well a contribution
that remains within the same Landau level), while not changing the value of $s\hbar\ell$ since the operators $(a_+,a^\dagger_+)$ act jointly
with $(a_-,a^\dagger_-)$ only through the combinations $a^\dagger_+ a^\dagger_-$ or $a_+ a_-$. Given (\ref{eq:L3}), it is straightforward
to spell out in detail the complete result for the matrix elements $\langle s\hbar\ell_1,E_{n_1}|\hat{L}_3| s\hbar\ell_2,E_{n_2}\rangle$, which is not
listed here. For identical values of $n_1=n_2$ however, one simply has,
\begin{equation}
\langle s\hbar\ell_1,E_{n_1}|\hat{L}_3| s\hbar\ell_2, E_{n_1}\rangle = - s\hbar (2n_1+1)\,\delta_{\ell_1,\ell_2}.
\end{equation}
Using the fact that
\begin{equation}
|T_1,E_{n_-}\rangle = \sum_{n_+=0}^\infty |n_+,n_-\rangle \langle n_+,n_-|T_1,E_{n_-}\rangle,
\end{equation}
and that the change of basis matrix elements $\langle n_+,n_-|T_1,E_{m_-}\rangle$ is diagonal in $(n_-,m_-)$,
it readily follows that,
\begin{equation}
\langle T_1,E_{n_1}|\hat{L}_3|T'_1,E_{n_1}\rangle = -s\hbar (2n_1+1)\,\delta(T_1-T'_1).
\end{equation}
These results, listed in Tables~\ref{Table1} and \ref{Table2}, are of course also consistent with the following commutation relations,
\begin{equation}
\left[\hat{T}_i,\hat{L}_3\right]=-i\hbar\,\epsilon_{ij}\,\hat{p}_j,\qquad
\left[\hat{p}_i,\hat{L}_3\right]=i\hbar\,\epsilon_{ij}\,\hat{T}_j\,-\,2 i\hbar\,\epsilon_{ij}\,\hat{p}_j,
\end{equation}
and their matrix elements.

As a final comment, let us point out that any of the gauge invariant and physical quantum observables considered herein are represented by matrix elements
which are of course different for different choices of eigenbases. Even though as abstract quantum operators they are of course independent of such a choice,
\textcolor{black}{as they are as well truly independent} from the choice of gauge for the vector potential representing the uniform and static magnetic field $\vec{B}$.
None of the matrix elements listed in Tables~\ref{Table1} and \ref{Table2} depend on the values of the general gauge fixing parameters $(\alpha,\varphi(\vec{x}\,))$.
All possible canonical quantum realisations of the Landau problem are unitarily equivalent, inclusive of all possible gauge transformations of the background
vector gauge potential representing the uniform and static magnetic field.

\subsection{\textcolor{black}{Assessing a recent publication dedicated to these issues}}
\label{Sect5.5}

\textcolor{black}{
Based on the considerations and results developed and achieved in the present work through detailed and careful analysis,
let us compare these with statements and general conclusions to be found in the recent publication of Ref.\cite{Waka1}.
While objections could be addressed to a series of other different elements of that work with their varying degrees of importance, here only the primary
issues with some statements and the main conclusion of Ref.\cite{Waka1} are pointed out explicitly.}

\textcolor{black}{
As emphasized repeatedly in Ref.\cite{Waka1}, which ambitions to clarify ``some perplexing confusion'' that exists in the quantum Landau problem literature,
its main and most important result at least in the eyes of its authors is their claim (see for instance their Conclusion) that for the quantum Landau problem
 ``[...] This means that three gauge-potential-independent extensions do actually belong to {\sl different
gauge classes}, and that the conserved momentum\footnote{\textcolor{black}{In our notations, namely the Noether charges $\hat{\vec{T}}$.}}
and conserved OAM\footnote{\textcolor{black}{Conserved OAM: conserved orbital angular-momentum. In our notations, namely the Noether charge $\hat{M}_3$.}}
are {\sl not} truly gauge-invariant physical quantities despite their covariant gauge-transformation property [...].'' ({\sl sic}) However as our analysis has
established beyond any doubt, when all the consequences of gauge transformations in the background vector potential are properly accounted for
given the considered system, a same and unique unitarily equivalent gauge invariant quantum physics follows from the canonical quantisation
of the Landau problem, irrespective of the choice of gauge fixing for its vector potential. There exists a single and unique U(1) gauge equivalence class
for the quantum Landau problem and its observable physics in the  unbounded plane. Ref.\cite{Waka1}'s main conclusion is misconceived.}

\textcolor{black}{But how did its authors come to their untenable conclusion? Through an argumentation based on considerations presented
(primarily) in the two paragraphs that follow Eq.(146) of Ref.\cite{Waka1} (as well as at some other places, such as after their Eq.(142))
regarding the values taken by the
following matrix elements of the conserved total angular-momentum within a same Landau level, namely in our notations,
\begin{eqnarray}
\langle T_1,E_n|\hat{M}_3|T'_1,E_n\rangle &=& -\frac{1}{2}s \hbar^2 m \omega_c\, \delta''(T_1-T'_1)
+s \hbar\left(\frac{T^2_1}{2\hbar m \omega_c}-(n+\frac{1}{2}\right) \delta(T_1 -T'_1), \nonumber \\
\langle s\hbar\ell_1,E_n|\hat{M}_3|s\hbar\ell_2,E_n\rangle &=& s \hbar \ell_1\,\delta_{\ell_1,\ell_2},
\end{eqnarray}
evaluated (as done in Ref.\cite{Waka1}) whether in the first Landau gauge $(\alpha=1,\varphi(\vec{x}\,)=0)$,
or in the symmetric gauge $(\alpha=0,\varphi(\vec{x}\,)=0)$ (the same points apply as well to the same matrix elements of the conserved total momentum
operator, namely in our notations $\hat{\vec{T}}$\,). Each of these two separate sets of matrix elements for the same gauge invariant thus physical abstract
quantum observable $\hat{M}_3$ and for each of these two ensembles of gauge invariant abstract quantum states have indeed values that are independent
of the choice of gauge fixing parameters $(\alpha,\varphi(\vec{x}\,))$, and are thus indeed gauge invariant. However, based on the observation that for a given
Landau level $n$ the values for the gauge invariant matrix elements $\langle T_1,E_n|\hat{M}_3|T'_1,E_n\rangle$ and
$\langle s\hbar\ell_1,E_n|\hat{M}_3|s\hbar\ell_2,E_n\rangle$ are explicitly different, Ref.\cite{Waka1} argues (see the paragraph after its Eq.(142), as
well as the two paragraphs after its Eq.(146)) that ``What is important to recognize here is that these three types of eigen-states belong to totally different
(or inequivalent) gauge classes'' ({\sl sic}), by advocating the following criterion, ``The expectation values of a genuinely {\sl gauge-invariant} physical quantity
should be the {\sl same} irrespectively of the choice of three types of eigen-functions." ({\sl sic}) (note that the three types of eigen-states being refereed to are
the abstract gauge invariant joint eigenstates $|T_1,E_n\rangle$, $|T_2,E_n\rangle$ and $|s\hbar\ell,E_n\rangle$ of the pairs of abstract
gauge invariant commuting Noether charges $(\hat{T}_1,\hat{H})$, $(\hat{T}_2,\hat{H})$ and $(\hat{M}_3,\hat{H})$, respectively). But most obviously
the simple and ready reason why these matrix elements take different values for different choices of eigenstates has no relation whatsoever with the issue of gauge
transformations and gauge invariance. It is most basically the simple fact that a same abstract linear operator acting on some abstract vector space is represented
relative to some choice of basis by matrix elements whose values change by changing the basis, yet representing the same abstract operator.
The conserved total angular-momentum $\hat{M}_3$ is without any doubt a genuinely gauge invariant physical quantum observable whose
representation in terms of matrix elements relative to any choice of basis, say $|T_1,E_n\rangle$, $|T_2,E_n\rangle$ or $|s\hbar \ell,E_n\rangle$,
will involve different values for these matrix elements given any other choice of basis. In particular, the two bases $|T_1,E_n\rangle$ and
$|s\hbar\ell,E_n\rangle$ (for instance) span a same abstract Hilbert space and are related through a specific unitary transformation
of which the matrix elements $\langle T_1,E_{n_1}|s\hbar\ell,E_{n_2}\rangle$ are known and presented in Section \ref{Sect5.3}, and as a matter
of fact are also given in Ref.\cite{Waka1}
(in exactly the same way that the position eigenbasis and the energy or Fock eigenbasis of the ordinary harmonic oscillator are mapped into each other
through some unitary transformation, leading to a single unitary equivalence class for the quantum oscillator). Hence all three types of gauge invariant
eigenstate bases belong to a single unitary gauge equivalence class for the quantum Landau problem and its abstract Hilbert space,
irrespective of the gauge fixing chosen for its background
vector potential. Thus, what is perplexing indeed is that Ref.\cite{Waka1} should advocate
the untenable criterion cited above as a characterisation of a genuine gauge invariant physical quantity, thereby leading as well to the main untenable claim
of that work.}

\textcolor{black}{
Ref.\cite{Waka1} repeatedly also makes the claim, as it does for instance in the first citation provided above, that (again in our notations) the operators
$\hat{\vec{T}}$, $\hat{\vec{p}}$, $\hat{M}_3$ and $\hat{L}_3$ transform {\sl covariantly} under a gauge transformation of the background vector potential.
However as has clearly been established (already in the general context of Section~\ref{Sect3}) through our analysis,
as abstract quantum operators these are strictly gauge invariant observables, not subjected to any transformation induced by a gauge transformation of the background vector potential. And this in precisely the same way that abstract quantum states are strictly invariant under such gauge transformations.
Nonetheless and as has repeatedly been pointed out in our discussion, the configuration space representations of abstract quantum states in terms
of their wave functions, as well as of abstract quantum operators in terms of the associated matrix elements---given a choice of configuration space
eigenbasis and of unitary representation of the abstract Heisenberg algebra---do transform covariantly or contravariantly under gauge transformations
of the background vector field, as discussed explicitly at the end of Section \ref{Sect3} as well as in the first part of Section \ref{Sect5}. And this solely
because of the induced {\sl passive} U(1) unitary transformation of the configuration space eigenbasis being used, and certainly not because of a
transformation of the abstract quantum states which indeed are strictly gauge invariant. In particular
when choosing to work with the canonical unitary configuration space representation of the Heisenberg algebra (with a vanishing U(1) flat connection)
for whatever gauge fixing choice for the background vector field of the Landau problem, as Ref.\cite{Waka1} does, under a gauge transformation
$\Delta\bar{\varphi}(\vec{x}\,)$ the configuration space eigenbasis $|\vec{x}\,\rangle$ transforms covariantly as
\begin{equation}
|\vec{x}; 2\rangle = e^{-\frac{i}{\hbar}q\Delta\bar{\varphi}(\vec{x}\,)}\,|\vec{x}\,\rangle,
\end{equation}
while the configuration space representation of any gauge invariant abstract quantum state $|\psi\rangle$ transforms contravariantly as
\begin{equation}
\psi_{(2)}(\vec{x}\,)=\langle\vec{x}; 2|\psi\rangle=e^{\frac{i}{\hbar}q\Delta\bar{\varphi}(\vec{x}\,)}\,\langle\vec{x}|\psi\rangle
=e^{\frac{i}{\hbar}q\Delta\bar{\varphi}(\vec{x}\,)}\,\psi(\vec{x}\,).
\label{eq:change2}
\end{equation}
Consequently when not maintaining the nonetheless certainly necessary distinction, on the one hand, between abstract quantum states and operators,
and on the other hand, to be distinguished from their configuration space representations respectively
in terms of functions and differential operators given a choice of unitary configuration space representation of the Heisenberg algebra---indeed Ref.\cite{Waka1}
directly equates quantum states with their wave functions in configuration space, and likewise for quantum operators and their matrix elements---, there
unavoidably arises confusion regarding the gauge invariance or covariance of states and operators. This observation provides the reason for the unfounded
statement of Ref.\cite{Waka1} that the abstract operators $\hat{\vec{T}}$, $\hat{\vec{p}}$, $\hat{M}_3$ and $\hat{L}_3$ are covariant under gauge transformations
of the background vector potential, whereas they are in fact strictly invariant under such transformations. And thus these abstract operators
are all as well perfectly acceptable
as gauge invariant physical quantum observables, representing distinct and complementary physical properties of the point charge of the quantum Landau problem.}

\textcolor{black}{
To emphasize this last point yet once again, let us recall the basic and simple reason why these abstract quantum operators are indeed strictly
gauge invariant. Consider the Landau problem for two distinct choices of the background vector potential $\vec{A}^{(\alpha)}(\vec{x}\,)$ $(\alpha=1,2)$
which are gauge equivalent under a gauge transformation $\varphi_{21}(\vec{x}\,)$ such that
\begin{equation}
\vec{A}^{(2)}(\vec{x}\,)=\vec{A}^{(1)}(\vec{x}\,)+\vec{\nabla}\varphi_{21}(\vec{x}\,).
\end{equation} 
The associated canonical conjugate momenta $\vec{\pi}[\vec{A}^{(\alpha)}]=m\dot{\vec{x}}+q\vec{A}^{(\alpha)}(\vec{x}\,)$ $(\alpha=1,2)$
are then gauge dependent and related as
\begin{equation}
\vec{\pi}[\vec{A}^{(2)}]=\vec{\pi}[\vec{A}^{(1)}]+q\vec{\nabla}\varphi_{21}(\vec{x}),
\end{equation}
thereby implying the strict gauge invariance of the following phase space combinations
\begin{equation}
\vec{\pi}[\vec{A}^{(2)}]-q\vec{A}^{(2)}(\vec{x}\,)=\vec{\pi}[\vec{A}^{(1)}]-q\vec{A}^{(1)}(\vec{x}\,),
\end{equation}
which precisely coincide with the velocity momentum $\vec{p}=m\dot{\vec{x}}=\vec{\pi}[\vec{A}^{(1)}]-q\vec{A}^{(1)}(\vec{x}\,)=\vec{\pi}[\vec{A}^{(2)}]-q\vec{A}^{(2)}(\vec{x}\,)$.
For the canonically quantised system, let us introduce the unitary quantum operator constructed from the gauge transformation of the background vector potential,
\begin{equation}
U_{21}(\hat{\vec{x}})=e^{\frac{i}{\hbar}q\varphi_{21}(\hat{\vec{x}})}.
\end{equation}
Because of the Heisenberg algebra commutation relations $\left[ \hat{x}_i,\hat{\pi}[\vec{A}^{(\alpha)}]_j \right] = i\hbar\,\delta_{ij}\,\mathbb{I}$,
the abstract canonical conjugate
momenta operators $\hat{\vec{\pi}}[\vec{A}^{(\alpha)}]$ thus transform covariantly under gauge transformations, since one readily finds
\begin{equation}
\hat{\vec{\pi}}[\vec{A}^{(2)}]=\hat{\vec{\pi}}[\vec{A}^{(1)}]+q\vec{\nabla}\varphi_{21}(\hat{\vec{x}})=
U^\dagger_{21}(\hat{\vec{x}})\,\hat{\vec{\pi}}[\vec{A}^{(1)}]\,U_{21}(\hat{\vec{x}}),
\end{equation}
while the configuration space abstract quantum degrees of freedom $\hat{\vec{x}}$ are certainly strictly gauge invariant,
with obviously $U^\dagger_{21}(\hat{\vec{x}})\,\hat{\vec{x}}\,U_{21}(\hat{\vec{x}})=\hat{\vec{x}}$. Furthermore concurrently with the gauge transformation
of the abstract phase space quantum operators $(\hat{\vec{x}},\hat{\vec{\pi}}[\vec{A}^{(\alpha)}])$, the gauge transformation of the background
vector potential must be effected as well. Consequently the specific combination which defines the abstract velocity momentum quantum operator,
$\hat{\vec{p}}=\hat{\vec{\pi}}[\vec{A}^{(\alpha)}]-q\vec{A}^{(\alpha)}(\hat{\vec{x}})$, possesses altogether the following transformation under the gauge transformation
of the external background vector field,
\begin{equation}
\hat{\vec{\pi}}[\vec{A}^{(2)}]-q\vec{A}^{(2)}(\hat{\vec{x}})=U^\dagger_{21}(\hat{\vec{x}})\,\hat{\vec{\pi}}[\vec{A}^{(1)}]\,U_{21}(\hat{\vec{x}})
-\left(\vec{A}^{(1)}(\hat{\vec{x}})+q\vec{\nabla}\varphi_{21}(\hat{\vec{x}})\right)=\hat{\vec{\pi}}[\vec{A}^{(1)}]-q\vec{A}^{(1)}(\hat{\vec{x}}).
\end{equation}
In other words, the abstract velocity momentum quantum operator $\hat{\vec{p}}$ is indeed strictly gauge invariant, as is of course this physical observable
for the classical system as well.
By extension, any other abstract quantum operator constructed specifically only out of the gauge invariant operators $(\hat{\vec{x}},\hat{\vec{p}}\,)$
(rather than $\hat{\vec{\pi}}[\vec{A}^{(\alpha)}]$ on its own) is necessarily also strictly
gauge invariant. Obviously, by referring back to their expressions in terms only of $(\hat{\vec{x}},\hat{\vec{p}}\,)$, this is the case in particular for all conserved
Noether charges $\hat{H}$, $\hat{\vec{T}}$ and $\hat{M}_3$, as well as for the nonconserved velocity momentum and orbital angular-momentum operators,
$\hat{\vec{p}}$ and $\hat{L}_3$, respectively. All these operators are thus indeed strictly gauge invariant under gauge transformations
of the external background vector field. Not a single one of them transforms covariantly under these transformations.}

\textcolor{black}{
Fortunately the confusion regarding the actual strict gauge invariance of the abstract quantum operators $\hat{\vec{T}}$, $\hat{\vec{p}}$, $\hat{M}_3$ and $\hat{L}_3$
does not affect the evaluations of matrix elements presented in Ref.\cite{Waka1}, for the simple reason that
the relation (\ref{eq:change2}) is taken for granted and used as such when Ref.\cite{Waka1} applies gauge transformations
to the choice of background vector potential.
Ref.\cite{Waka1} considers matrix elements only for quantum states that belong to a same Landau level $n$, while the matrix elements
listed above in Tables~\ref{Table1} and \ref{Table2} apply to any pair of Landau levels $n_1$ and $n_2$. However these Tables do not include
the gauge dependent matrix elements for the gauge variant and thus not physically observable canonical operators $\hat{\vec{\pi}}[\vec{A}\,]$
and $\hat{L}^c_3[\vec{A}\,]$, while Ref.\cite{Waka1} considers these for the first Landau gauge and for the symmetric gauge.
Yet for all matrix elements that coincide, these two sets of independent evaluations do indeed agree.}

\textcolor{black}{
In particular Ref.\cite{Waka1} notes that the matrix elements (in our notations) $\langle T_1,E_n|\hat{p}_1|T'_1,E_n\rangle$ and
$\langle s\hbar\ell_1,E_n|\hat{p}_1|s\hbar\ell_2,E_n\rangle$ all vanish identically (this also applies to $\hat{p}_2$; see Tables \ref{Table1} and \ref{Table2}),
and then offers a tentative physics justification for such a result in terms of a time average of the classical cyclotron motion.
However this tentative justification is unfounded. On the one hand it is only for a pair of states
belonging to a same Landau level $n$ that all these matrix elements vanish identically. While on the other hand, because the velocity momentum operator
$\hat{\vec{p}}$ does not commute with $\hat{H}$, is not conserved and maps between adjacent Landau levels, for a pair of states belonging
to successive Landau levels none of the matrix elements of the velocity momentum operator $\hat{\vec{p}}$ vanish (see Tables \ref{Table1} and \ref{Table2}).
This simple property of $\hat{\vec{p}}$ is the sole reason why all its matrix elements for any pair of states belonging to a same Landau level vanish identically.
As a quantum operator acting on the full Hilbert space of Landau levels, $\hat{\vec{p}}$ certainly does not vanish.}

\textcolor{black}{
This same observation of vanishing matrix elements
also prompts Ref.\cite{Waka1} to argue that while the velocity momentum operator $\hat{\vec{p}}$ is physical in character and its
probability distribution may be observed, in contradistinction and even though gauge invariant as well, the conserved total momentum operator
and Noether charge $\hat{\vec{T}}$ does not share such a character and does not offer
``a direct connection with observables of the Landau electron'' ({\sl sic}). However as has been established earlier, up to normalisation the dual components
of $\hat{\vec{T}}$ are essentially the components of the conserved gauge invariant magnetic centre quantum operator $\hat{\vec{x}}_c$, which no doubt is certainly
also a physically observable quantity for the point charge of the quantum Landau problem. Hence so is certainly the Noether charge $\hat{\vec{T}}$ as well,
as the generator of translations in the plane. As is as well the Noether charge $\hat{M}_3$ measuring the conserved total angular-momentum of the system,
as generator of rotations in the plane centered at the arbitrarily chosen point $\vec{x}_0$.}

\textcolor{black}{
As a matter of fact, it has been shown in the first part of Section \ref{Sect5} that the gauge invariant vector operators $\hat{\vec{p}}$ and $\hat{\vec{T}}$,
or equivalently $\hat{\vec{p}}$ and $\hat{\vec{x}}_c$, commute with each other (but not their respective cartesian components among themselves). Hence
any one cartesian component of each would provide a pair of a maximal set of commuting gauge invariant physical observables whose joint eigenbasis
would define yet another possible basis of the same Hilbert space of quantum states of the quantum Landau problem, even though $\hat{\vec{p}}$ is not conserved.}

\textcolor{black}{
Further issues with other statements and considerations developed in Ref.\cite{Waka1} could be raised as well, but will not be included here
and are left for the reader to assess.}

\section{Conclusions}
\label{Sect6}

Any coupled set of classical equations of motion that derives from an action principle is left invariant under any \textcolor{black}{redefinition}
of that action which simply adds
to it an arbitrary surface or divergence term, namely a total time derivative surface term in the case of a mechanical system (or a total space-time
surface term in the case of a field theory).

Even though leaving the classical equations of motion invariant, for the classical dynamics in its canonical Hamiltonian formulation
such a redefinition of the classical action induces a nontrivial canonical transformation in phase space leading to transformed conjugate momenta
given the invariant original configuration space degrees of freedom, yet with an identical canonical Hamiltonian.

For the canonically quantised dynamics,
such a redefinition of the classical action induces not only the canonical phase space transformation now realised as a unitary transformation
acting on the conjugate momenta as abstract quantum operators, but effects concurrently a transformation in the unitary representation of the abstract
Heisenberg algebra that is being used for the configuration space wave function representation of abstract quantum states in Hilbert space.
\textcolor{black}{Up to unitary transformations,} these unitary \textcolor{black}{configuration space}
representations are \textcolor{black}{organised into} U(1) gauge equivalence classes of flat U(1) connections over configuration space, namely
the U(1) representations of the first homotopy group of the configuration space manifold, characterised by the U(1) holonomies for all noncontractible cycles
in that manifold \cite{Gov2}. Redefining the classical action by a surface term induces for the canonically quantised system a U(1) local gauge transformation
of the initially chosen representative flat U(1) connection that defines a specific unitary representation of the Heisenberg algebra within the same U(1) gauge equivalence class of such representations. Furthermore this transformation in the unitary representation of the Heisenberg algebra is correlated with possible changes
in the local quantum phases of the configuration space eigenstates used to represent abstract quantum states in terms of configuration space wave functions.
When these specific features are properly accounted for in the canonically quantised system, the resulting quantum realisation remains indeed
physically identical for all abstract quantum observables and all abstract quantum states, irrespective of the specific surface term by which the classical action
is \textcolor{black}{redefined} or not.

Such subtle quantum effects resulting from an apparently trivial redefinition of a classical action which leaves its classical equations of motion invariant
must certainly be considered when the canonically quantised system is coupled to some classical external background gauge field.
Indeed under any gauge transformation
{\it of the background gauge field, while the configuration space degrees of freedom are \textcolor{black}{strictly} left invariant},
the action then transforms precisely by an additive
surface term.

A proper and complete understanding of the consequences of gauge transformations of {\it the background gauge field \textcolor{black}{on its own}}, especially
for the canonically quantised system and \textcolor{black}{the gauge invariance} for all its physical abstract quantum observables and abstract quantum states,
requires a careful consideration of the quantum transformations induced in Hilbert space and for its abstract operator algebra, and their unitary representations,
by the gauge transformation of the background gauge field itself. These different transformations must be considered jointly, indeed to ascertain once again
that the resulting quantum realisation of the system coupled to the background gauge field is physically identical for all physical and gauge invariant
abstract quantum observables and abstract quantum states, irrespective of gauge transformations of the background gauge field, even though
configuration space wave function representations \textcolor{black}{themselves} of \textcolor{black}{abstract gauge invariant} quantum states may vary
(through U(1) phase factors) as the background gauge field itself is being gauge transformed.

After a detailed discussion of these different considerations in a general context, their conclusions have been illustrated explicitly in detail
and with care in the case of the celebrated
Landau problem and its canonical quantisation, by also emphasizing in that case the important role played by the continuous and global space-time symmetries
of that specific two dimensional system in the Euclidean plane of a nonrelativistic point charge coupled to a uniform and static transverse background magnetic field.

The present revisit of a most studied system involves a parametrisation of the choice of gauge for the background magnetic vector potential that is as general and
as comprehensive as possible, without restricting to any one of the three specific gauge choices considered time and again in the literature, which each proves
to be such that one of the gauge variant conjugate momenta (whether in cartesian or polar circular coordinates) then coincides with one of the gauge invariant
and physical conserved Noether charges \textcolor{black}{(other than the generator for time evolution)} that generate the system's global space-time symmetries.

A careful analysis has been presented, paying due attention to the
subtleties relating to gauge transformations of the background vector gauge potential, the use of unitary representations of the Heisenberg algebra, and the
choice of local quantum phases of configuration space eigenstates to construct wave functions representations of the abstract quantum states that diagonalise
the gauge invariant and physical quantum Hamiltonian jointly with one of the three gauge invariant and physical conserved Noether charges generating the spatial
global symmetries of the quantum dynamics. The role of the choice of gauge for the vector potential is thereby made transparent and explicit at each stage
of the analysis, and how its gauge transformations induce the U(1) transformations of the quantised dynamics as they are indeed necessary to ensure
gauge invariance of the intrinsically physical abstract quantum observables and quantum states, inclusive of their quantum wave functions and
matrix elements.

Results established herein \textcolor{black}{ specifically for a series of matrix elements of gauge invariant and physical abstract quantum observables}
compare favourably with those provided in Ref.\cite{Waka1}. However a number of considerations developed, and
conclusions drawn in the present work through a careful analysis of these more subtle points having to do in particular with transformations in unitary
representations of the Heisenberg algebra of the canonically quantised system under gauge transformations of the classical external field,
certainly do not support a series of statements, comments and (some strong) claims made in that recent publication.

The discussions detailed herein aim to clarify some confusion \textcolor{black}{that exists}
within the literature surrounding the issue of the choice of gauge and of gauge transformations of the classical external background vector potential
in the quantum Landau problem and its quantum dynamics.

While precisely the same general considerations of the subtle interplay between gauge transformations of the external \textcolor{black}{classical}
gauge field and the induced transformations
for the canonically quantised system coupled to it, especially for the transformation in the unitary representation of the Heisenberg algebra being used,
must be addressed whenever a quantum dynamics is coupled to a classical external background gauge field, whatever the geometry of its configuration space
which itself is certainly not transformed under gauge transformations of the background gauge field.

\section*{Acknowledgement}

This work is supported in part by the {\sl Institut Interuniversitaire des Sciences Nucl\'eaires} (IISN, Belgium).

\end{document}